\documentclass[review,12pt,letterpaper]{elsarticle}

\usepackage[utf8]{inputenc}
\usepackage[T1]{fontenc}
\usepackage{graphicx}
\usepackage{mathrsfs}
\usepackage[intlimits,centertags]{amsmath}
\usepackage{amssymb,amsfonts}
\usepackage[pdftex]{hyperref}
\usepackage[x11names]{xcolor}
\usepackage{aas_macros}
\usepackage{ulem}

\hypersetup{pdftitle={Opening a New Window onto the Universe with IceCube},
pdfsubject={Opening a New Window onto the Universe with IceCube},
pdfauthor={Markus Ahlers and Francis Halzen},
pdfstartview={FitH},
colorlinks=true,
bookmarksopen=false,
bookmarksnumbered=false,
bookmarksopenlevel=0,
linkcolor=Blue1!60!black,
citecolor=Green1!50!black,
urlcolor=Blue1!70!black
}

\journal{Progress in Particle and Nuclear Physics}

\begin{document}

\begin{frontmatter}

\title{\vspace*{1cm}Opening a New Window onto the Universe with IceCube}

\author{Markus Ahlers}
\address{Niels Bohr International Academy \& Discovery Centre, Niels Bohr Institute,\\ University of Copenhagen, DK-2100 Copenhagen, Denmark}

\author{Francis Halzen}
\address{Wisconsin IceCube Particle Astrophysics Center \& Department of Physics,\\ University of Wisconsin--Madison, Madison, WI 53706, USA}

\begin{abstract}
Weakly interacting neutrinos are ideal astronomical messengers because they travel through space without deflection by magnetic fields and, essentially, without absorption. Their weak interaction also makes them notoriously difficult to detect, with observation of high-energy neutrinos from distant sources requiring kilometer-scale detectors. The IceCube project transformed a cubic kilometer of natural Antarctic ice at the geographic South Pole into a Cherenkov detector. It discovered a flux of cosmic neutrinos in the energy range from 10~TeV to 10~PeV, predominantly extragalactic in origin.  Their corresponding energy density is close to that of high-energy photons detected by gamma-ray satellites and ultra-high-energy cosmic rays observed with large surface detectors. Neutrinos are therefore ubiquitous in the nonthermal universe, suggesting a more significant role of protons (nuclei) relative to electrons than previously anticipated. Thus, anticipating an essential role for multimessenger astronomy, IceCube is planning significant upgrades of the present instrument as well as a next-generation detector. Similar detectors are under construction in the Mediterranean Sea and Lake Baikal.
\end{abstract}


\end{frontmatter}

\section{Neutrino Astronomy}

Soon after the 1956 observation of the neutrino~\cite{Cowan:1992xc}, the idea emerged that it represented the ideal astronomical messenger. Neutrinos reach us from the edge of the universe without absorption and with no deflection by magnetic fields. Having essentially no mass and no electric charge, the neutrino is similar to the photon as an astronomical messenger, except for one important attribute: its interactions with matter are extremely feeble. 
Whereas high-energy photons (gamma rays) above 10~TeV energy are strongly absorbed by interactions with the cosmic microwave background (CMB), the universe remains transparent to neutrinos across the spectrum. We know that radiation from the extreme universe extends to at least 100\,EeV because cosmic rays with this energy have been observed. However, the arrival directions of cosmic rays below 10~EeV are strongly distorted by deflections in cosmic magnetic fields. Therefore, six out of 29 decades in energy, ranging from low-energy (MHz) radiowaves to the highest energy cosmic rays, can only be covered by neutrino astronomy.

Originating from the decay of pions and kaons that are exclusively produced in sites where protons and nuclei are accelerated, neutrinos trace cosmic ray accelerators. Unfortunately, their weak interactions also make cosmic neutrinos very difficult to detect. It was already apparent in the early 1970s that immense particle detectors are required to collect cosmic neutrinos in statistically significant numbers~\cite{Roberts:1992re}.

Following a suggestion by Markov~\cite{Markov:1960vja}, early efforts concentrated on transforming a large volume of deep natural water instrumented with light sensors into a Cherenkov detector that collects the light produced when neutrinos interact with nuclei in or near the instrumented volume. After a two-decade-long effort, building the Deep Underwater Muon and Neutrino Detector (DUMAND) in the sea off the main island of Hawaii unfortunately failed~\cite{Babson:1989yy}. However, DUMAND paved the way for later efforts by pioneering many of the detector technologies in use today and by inspiring the deployment of a smaller instrument in Lake Baikal~\cite{Balkanov:2000cf}. A similar instrument, ANTARES, has been operating in the Mediterranean Sea for over a decade~\cite{Aggouras:2005bg,Aguilar:2006rm,Migneco:2008zz}, leading the way for the ongoing construction of KM3NeT~\cite{Adrian-Martinez:2016fdl} off the coasts of Sicily (ARCA) and France (ORCA) and GVD~\cite{Avrorin:2015wba} in Lake Baikal.

The first telescope on the scale envisaged by the DUMAND collaboration was realized by the turn of the century by transforming a large volume of transparent natural Antarctic ice into a particle detector, the Antarctic Muon and Neutrino Detector Array (AMANDA)~\cite{Andres:1999hm}. It represented a proof of concept for the kilometer-scale neutrino observatory, IceCube~\cite{ICPDD2001,Aartsen:2016nxy}. In 2013, two years after completing construction, IceCube announced the discovery of a flux of cosmic neutrinos with energies in the 30\,TeV--1\,PeV range. The large extragalactic flux observed implies that the energy density of neutrinos in the high-energy universe matches that observed in photons, indicating a much larger role of protons relative to electrons than anticipated. Indeed, a recent observation of a gamma-ray flare of a blazar, an active galaxy with jets aligned with the Earth, in spatial and temporal coincidence with a high-energy neutrino observed by IceCube may be the first evidence of an extragalactic cosmic ray source. We will summarize the status of neutrino astronomy further on in this review. 

In the following, we will give an update of the present status of neutrino astronomy~\cite{Ahlers:2015lln,Ahlers:2017wkk}. We will summarize recent neutrino observations with IceCube and highlight their context in the larger view of multimessenger astronomy.

\subsection{Detecting Very High-Energy Neutrinos}

Cosmic rays have been studied for more than a century. They reach energies in excess of $10^8$\,TeV, populating an extreme universe that is opaque to electromagnetic radiation. We do not yet know where or how particles are accelerated to these extreme energies, and with the recent observation of a blazar in coincidence with the direction and time of a very high energy muon neutrino, neutrino astronomy might have taken a first step in solving this puzzle~\cite{Kotera:2011cp,Ahlers:2015lln}. The rationale is simple: near neutron stars and black holes, gravitational energy released in the accretion of matter or binary mergers can power the acceleration of protons ($p$) or heavier nuclei that subsequently interact with gas (``$pp$'') or ambient radiation (``$p\gamma$''). Neutrinos are produced by cosmic-ray interactions at various places: in their sources during their acceleration, in the source periphery just after their release, and even in cosmic voids while propagating through universal radiation backgrounds.  In interactions of cosmic-ray protons with background photons ($\gamma_{\rm bg}$), neutral and charged pion secondaries are produced in the processes $p+\gamma_{\rm bg}\to p+\pi^0$ and $p+\gamma_{\rm bg}\to n+\pi^+$. While neutral pions decay as $\pi^0\to\gamma+\gamma$ and create a flux of high-energy gamma rays, the charged pions decay into three high-energy neutrinos ($\nu$) and anti-neutrinos ($\bar\nu$) via the decay chain $\pi^+\to\mu^++\nu_\mu$ followed by $\mu^+\to e^++\bar\nu_\mu +\nu_e$, and the charged-conjugate process. We refer to these photons as pionic photons to distinguish them from photons radiated by electrons that may be accelerated along with the protons and nuclei. Once produced, neutrinos do not take part in further interactions, unless produced within very dense environments. Due to their weak interaction, they essentially act like photons; their small mass is negligible relative to the TeV to EeV energies targeted by neutrino telescopes. They do however oscillate over cosmic distances. For instance, for an initial neutrino flavor ratio of  $\nu_e:\nu_\mu:\nu_\tau \simeq 1:2:0$ from the decay of pions and muons, the oscillation-averaged composition arriving at the detector is approximately an equal mix of electron, muon, and tau neutrino flavors, $\nu_e:\nu_\mu:\nu_\tau \simeq 1:1:1$~\cite{Farzan:2008eg}.

\begin{figure}[t]
\centering
\includegraphics[width=1.0\linewidth]{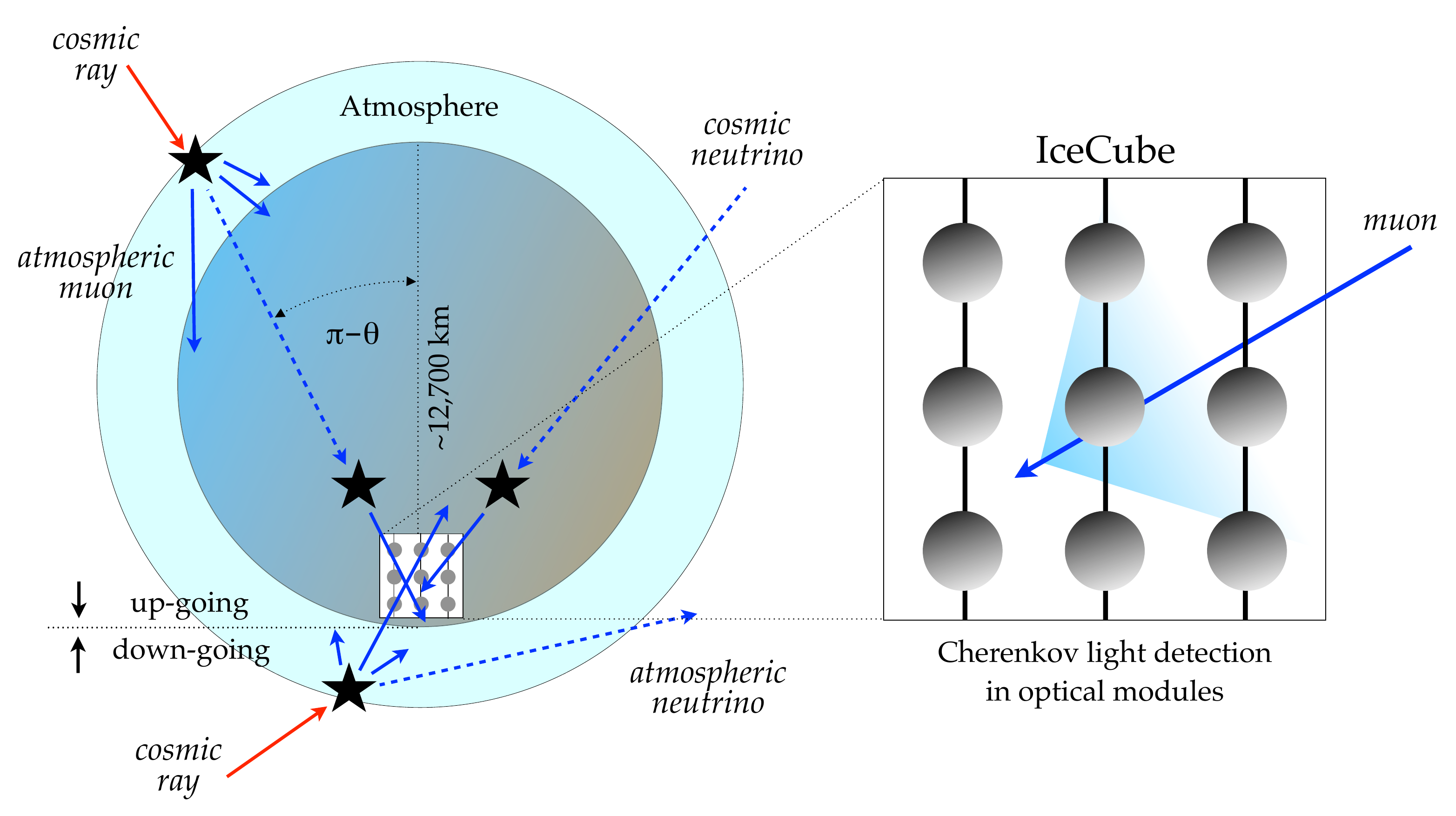}
\caption[]{The principal idea of neutrino telescopes from the point of view of IceCube located at the South Pole. Neutrinos dominantly interact with a nucleus in a transparent medium like water or ice and produce a muon that is detected by the wake of Cherenkov photons it leaves inside the detector. The background of high-energy muons (solid blue arrows) produced in the atmosphere can be reduced by placing the detector underground. The surviving fraction of muons is further reduced by looking for upgoing muon tracks that originate from muon neutrinos (dashed blue arrows) interacting close to the detector. This still leaves the contribution of muons generated by atmospheric muon neutrino interactions. This contribution can be separated from the diffuse cosmic neutrino emission by an analysis of the combined neutrino spectrum.}
\label{fig:earth}
\end{figure}

High-energy neutrinos interact predominantly with matter via deep inelastic scattering off nucleons: the neutrino scatters off quarks in the target nucleus by the exchange of a $Z$ or $W$ weak boson, referred to as {\it neutral current} (NC) and {\it charged current} (CC) interactions, respectively. Whereas the NC interaction leaves the neutrino state intact, in a CC interaction a charged lepton is produced that shares the initial neutrino flavor. The average relative energy fraction transferred from the neutrino to the lepton is at the level of $80$\% at these energies. The inelastic CC cross section on protons is at the level of $10^{-33}~{\rm cm}^{2}$ at a neutrino energy of $10^3$~TeV and grows with neutrino energy as $\sigma_{\rm tot}\propto E_\nu^{0.36}$~\cite{Gandhi:1995tf,CooperSarkar:2011pa}. The struck nucleus does not remain intact and its high-energy fragments typically initiate hadronic showers in the target medium.

Immense particle detectors are required to collect cosmic neutrinos in statistically significant numbers. Already by the 1970s, it had been understood~\cite{Roberts:1992re} that a kilometer-scale detector was needed to observe the cosmogenic neutrinos produced in the interactions of CRs with background microwave photons~\cite{Beresinsky:1969qj}. A variety of methods are used to detect the high-energy secondary particles created in CC and NC neutrino interactions. One particularly effective method observes the radiation of optical Cherenkov light given off by secondary charged particles produced in CC and NC interactions that travel faster than the speed of light in the medium. The detection concept is that of a Cherenkov detector, a transparent medium instrumented with photomultipliers that transform the Cherenkov light into electrical signals using the photoelectric effect; see Figs.~\ref{fig:earth} and \ref{fig:deepcore}. Computers at the surface use this information to reconstruct the light patterns produced in neutrino events and infer their arrival directions, their energies, and their flavor.

There are two principle classes of Cherenkov events that can be easily identified this way, ``tracks'' and ``cascades.'' The term ``tracks'' refers to the Cherenkov emission of long-lived muons passing through the detector. These muons can be produced in CC interactions of muon neutrinos inside or in the vicinity of the detector. Energetic electrons and taus produced in CC interactions of electron and tau neutrino interactions, respectively, will in general not produce elongated tracks due to the rapid scattering of electrons and the short lifetime of the tau. Because of the large background of muons produced by CR interactions in the atmosphere, the observation of muon neutrinos is limited to upgoing muon tracks that are produced in interactions inside or close to the detector by neutrinos that have passed through the Earth\footnote{Note, that while at high-energy the neutrino cross section grows, resulting in a reduced mean free path ($\lambda_\nu$), the range of the secondary muon ($\lambda_{\mu}$) increases as does the probability for observing a muon, $\lambda_\mu/\lambda_\nu$; it is about $10^{-6}$ for a 1~TeV neutrino.} as illustrated in Fig.~\ref{fig:earth}. The remaining background consists of atmospheric neutrinos, which are indistinguishable from cosmic neutrinos on an event-by-event basis. However, the steeply falling spectrum ($\propto E^{-3.7}$) of atmospheric neutrinos allows identifying diffuse astrophysical neutrino emission above a few hundred TeV by a spectral analysis, as we will highlight in the following sections. The atmospheric background is also reduced for muon neutrino observation from point-like sources, in particular transient neutrino sources.

The hadronic particle shower generated by the target struck by a neutrino in the ice also radiates Cherenkov photons. Because of the large multiplicity of secondary particles and the repeated scattering of the Cherenkov photons in the medium, the light pattern is mostly spherical; it is referred to as a  ``cascade.'' The light patterns produced by the particle showers initiated by the electron or tau produced in CC interactions of electron or tau neutrinos, respectively, will be superimposed on the cascade. The direction of the initial neutrino can only be reconstructed from the Cherenkov emission of secondary particles produced close to the neutrino interaction point, and the angular resolution is worse than for track events. 

On the other hand, the energy of the initial neutrino can be constructed with a better resolution than for tracks. For both tracks and cascades, the observable energy of secondary charged leptons can be estimated from the total number of Cherenkov photons and is related to the neutrino energy of charged particles after accounting for kinematic effects and detection efficiencies. The Cherenkov light observed in cascades is proportional to the energy transferred to the cascade and can be fully contained in the instrumented volume. In contrast, muons produced by CC muon neutrino interactions lose energy gradually by ionization, bremsstrahlung, pair production, and photo-nuclear interactions while they range in. This allows estimating the muon energy as it enters the detector and setting a lower limit on the neutrino energy.

\subsection{IceCube Neutrino Observatory}
\label{sec:IceCube}

\begin{figure}[t]\centering
\includegraphics[width=1.0\linewidth]{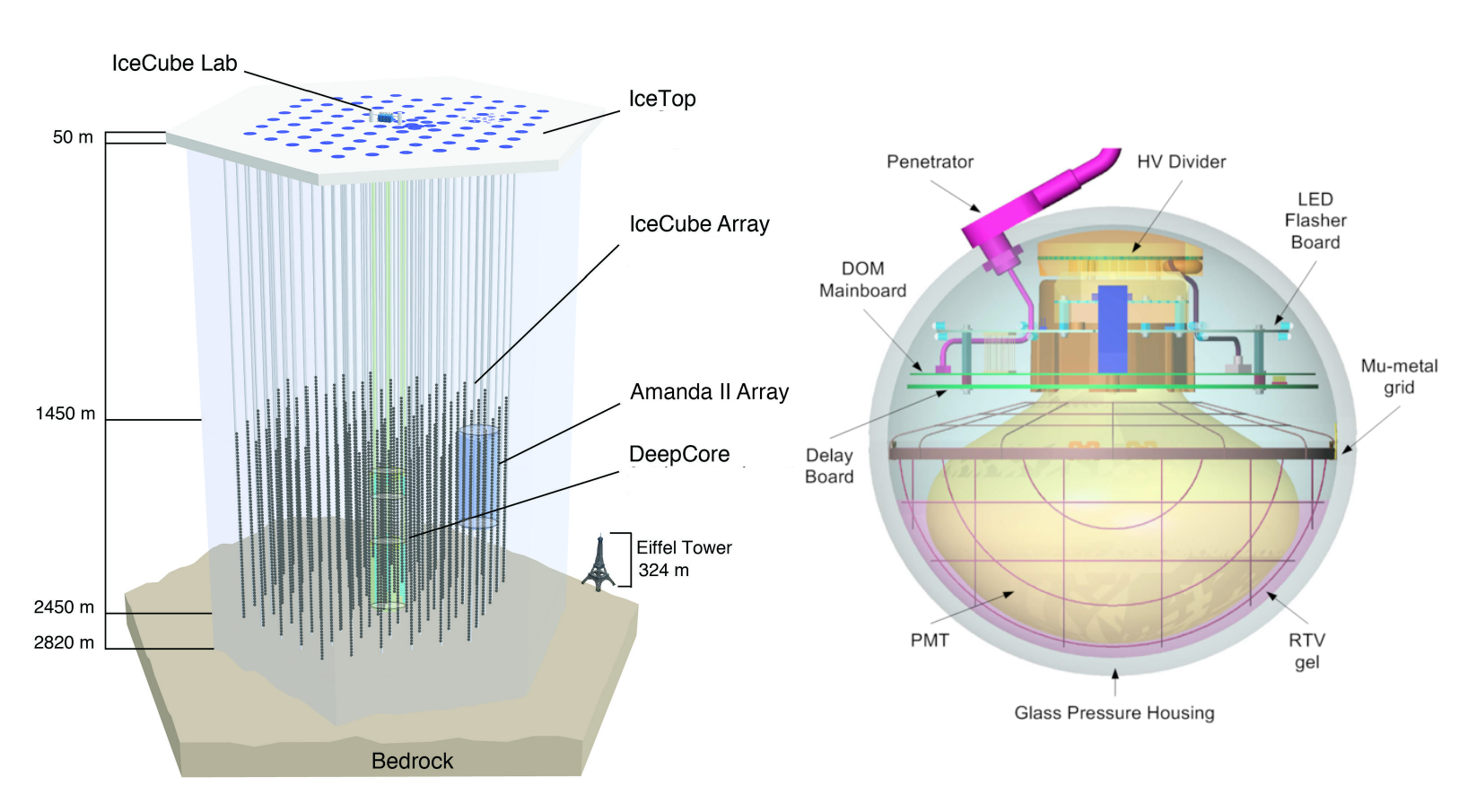}
\caption[]{Architecture of the IceCube observatory (left) and the schematics of a digital optical module (right) (see Ref.~\cite{Abbasi:2008aa} for details).}
\label{fig:deepcore}
\end{figure}

The IceCube detector~\cite{Aartsen:2016nxy} transforms the deep natural Antarctic ice 1,450 m below the geographic South Pole into a Cherenkov detector. A cubic kilometer of ice is instrumented with 5,160 digital optical modules; see Fig.~\ref{fig:deepcore}. Each digital optical module consists of a glass sphere that contains a 10-inch photomultiplier and an electronics board that digitizes the signals locally. The digitized signals are given a global time stamp with an accuracy of two nanoseconds and are subsequently transmitted to the surface. Processors at the surface continuously collect the time-stamped signals from the optical modules, each of which functions independently. These signals are sorted into telltale patterns of light that reveal the direction, energy, and flavor of the incident neutrino.

\begin{figure}[t]\centering
\includegraphics[width=0.43\linewidth,viewport=20 0 200 170,clip=true]{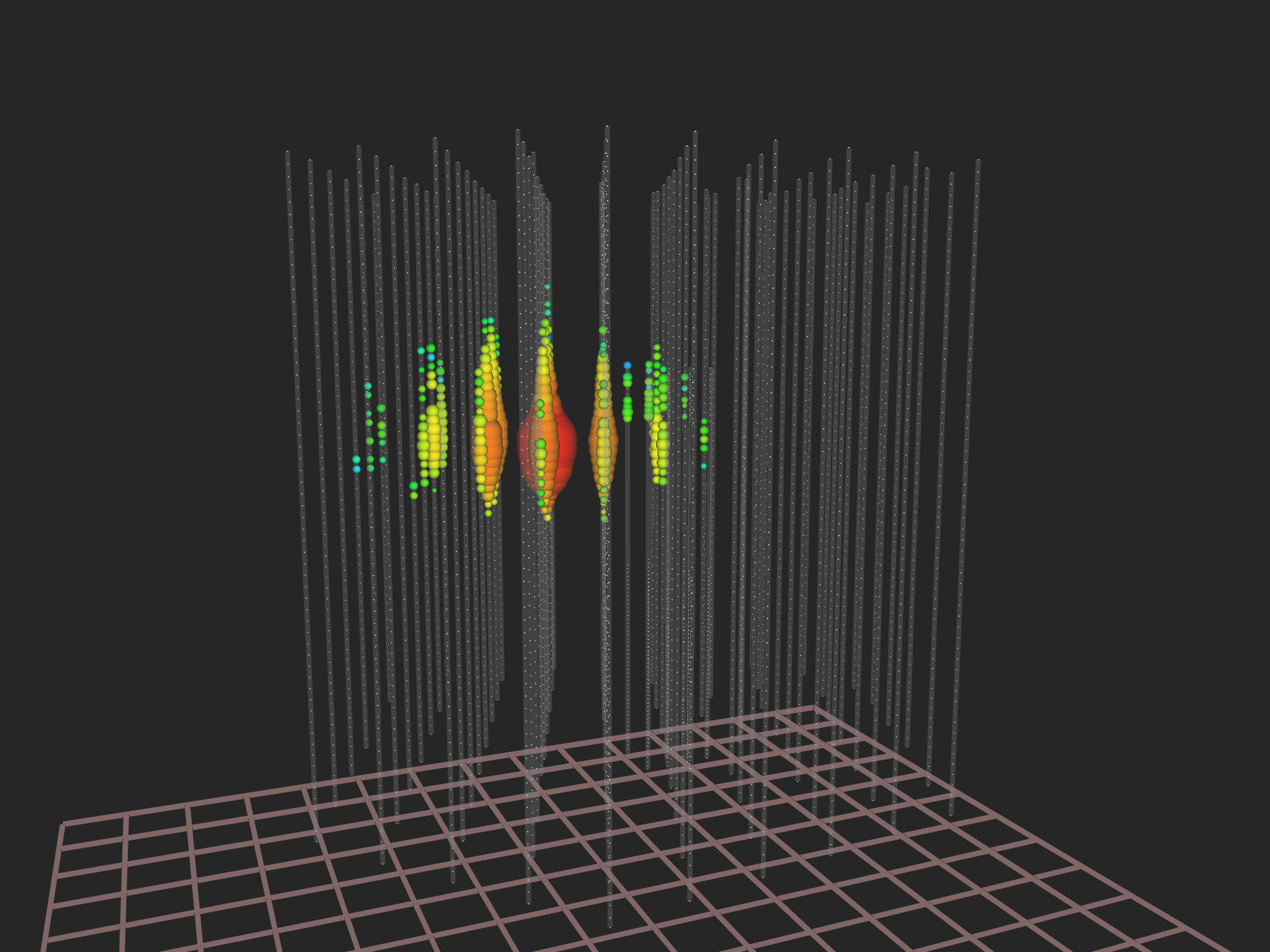}\hspace{0.5cm}\includegraphics[width=0.43\linewidth,viewport=20 0 200 170,clip=true]{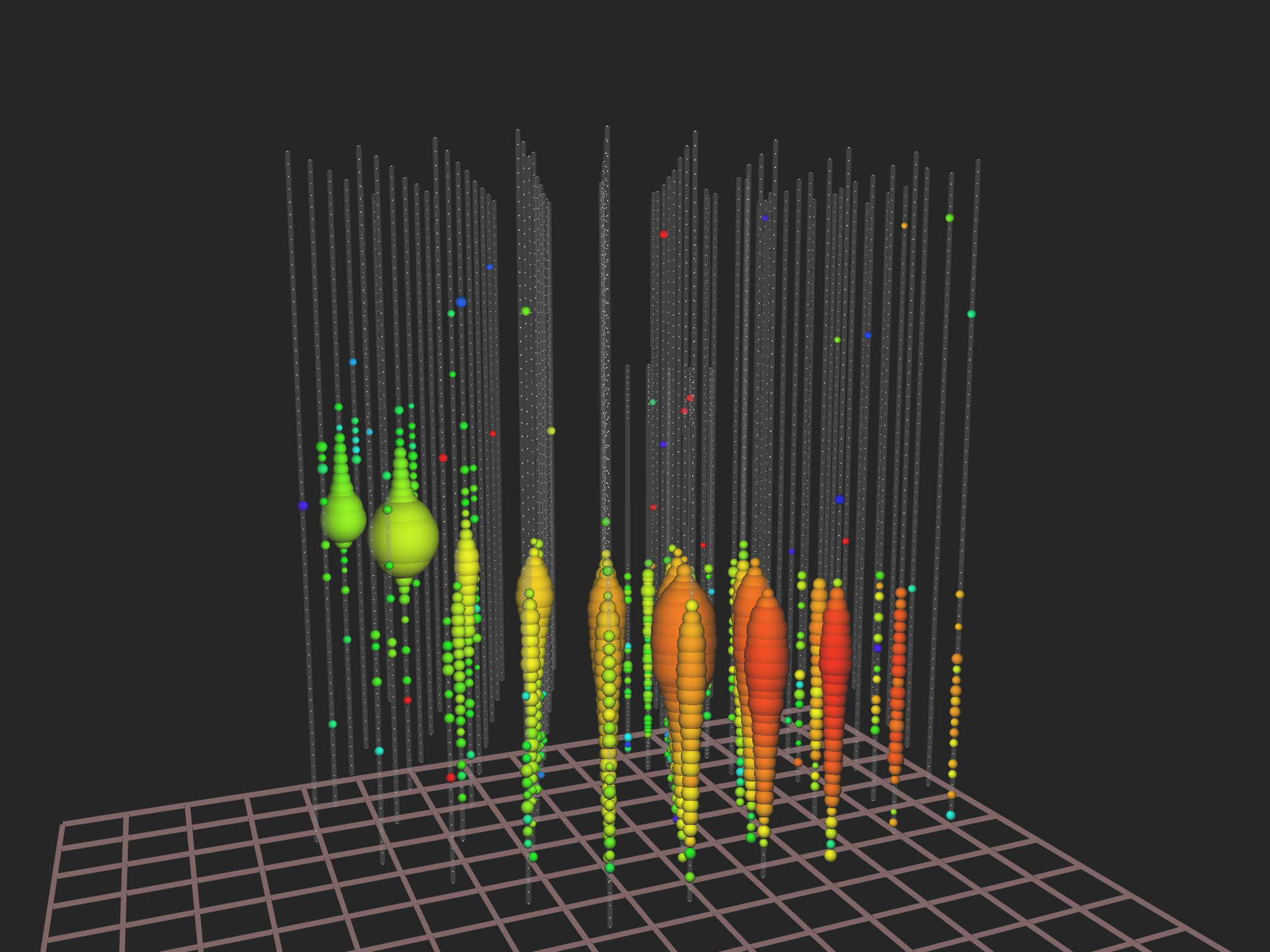}
\caption[]{{\bf Left Panel:}  Light pool produced in IceCube by a shower initiated by an electron or tau neutrino. The measured energy is $1.14$ PeV, which represents a lower limit on the energy of the neutrino that initiated the shower. White dots represent sensors with no signal. For the colored dots, the color indicates arrival time, from red (early) to purple (late) following the rainbow, and size reflects the number of photons detected. {\bf Right Panel:}  An upgoing muon track traverses the detector at an angle of $11^\circ$ below the horizon. The deposited energy, i.e., the energy equivalent of the total Cherenkov light of all charged secondary particles inside the detector, is 2.6\,PeV.}
\label{fig:erniekloppo}
\end{figure}

Because of its size, even at a depth of 1,450 m, IceCube detects a background of atmospheric cosmic-ray muons originating in the Southern Hemisphere at a rate of 3,000 per second. Two methods are used to identify neutrinos. Traditionally, neutrino searches have focused on the observation of muon neutrinos that interact primarily outside the detector to produce kilometer-long muon tracks passing through the instrumented volume. Although this allows the identification of neutrinos that interact outside the detector, it is necessary to use the Earth as a filter in order to remove the huge background of cosmic-ray muons. This limits the neutrino view to a single flavor and half the sky. An alternative method exclusively identifies high-energy neutrinos interacting inside the detector, so-called high-energy starting events (HESE). It divides the instrumented volume of ice into an outer veto shield and a $\sim420$-megaton inner fiducial volume. The advantage of focusing on neutrinos interacting inside the instrumented volume of ice is that the detector functions as a total absorption calorimeter, measuring the neutrino energy of cascades with a 10--15\,\% resolution~\cite{Aartsen:2013vja}. Furthermore, with this method, neutrinos from all directions in the sky can be identified, including both muon tracks as well as secondary showers, produced by charged-current interactions of electron and tau neutrinos, and neutral current interactions of neutrinos of all flavors. For illustration, the Cherenkov patterns initiated by an electron (or tau) neutrino of about~1\,PeV energy and a muon neutrino losing 2.6\,PeV energy in the form of Cherenkov photons while traversing the detector are contrasted in Fig.~\ref{fig:erniekloppo}.

In general, the arrival times of photons at the optical sensors, whose positions are known, determine the particle's trajectory, while the number of photons is a proxy for the deposited energy, i.e., the energy of all charged secondary particles from the interactions. For instance, for the cascade event shown in the left panel of Fig.~\ref{fig:erniekloppo}, more than 300 digital optical modules (DOMs) report a total of more than 100,000 photons. The two abovementioned methods of separating neutrinos from the cosmic-ray muon background have complementary advantages. The long tracks produced by muon neutrinos can be pointed back to their sources with a $\le 0.4^\circ$ angular resolution.  In contrast, the reconstruction of the direction of cascades in the HESE analysis, in principle possible to a few degrees, is still in the development stage in IceCube~\cite{Aartsen:2013vja}. They can be reconstructed to within $10^\circ\sim15^\circ$ of the direction of the incident neutrino. Determining the deposited energy from the observed light pool is, however, relatively straightforward, and a resolution of better than 15\,\% is possible; the same value holds for the reconstruction of the energy deposited by a muon track inside the detector.

\begin{figure}[t]\centering
\includegraphics[width=0.5\linewidth]{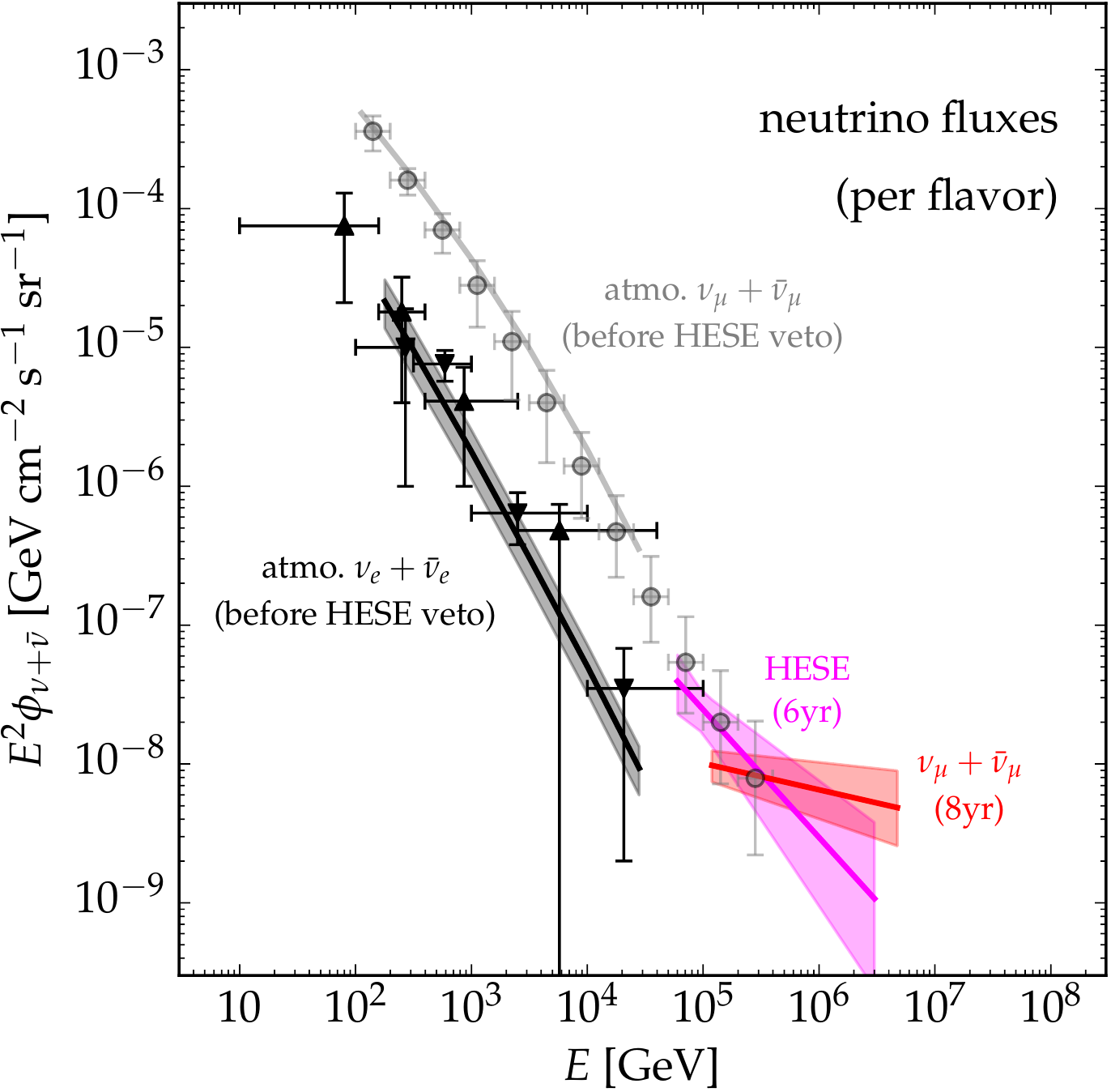}
\caption[]{Summary of diffuse neutrino observations (per flavor) by IceCube. The black and gray data show IceCube's measurement of the atmospheric $\nu_e+\bar\nu_e$~\cite{Aartsen:2012uu,Aartsen:2015xup} and $\nu_\mu +\bar\nu_\mu$~\cite{Abbasi:2010ie} spectra.  The magenta line and magenta-shaded area indicate the best-fit and $1\sigma$ uncertainty range of a power-law fit to the six-year HESE data. Note that the HESE analysis vetoes atmospheric neutrinos and can probe astrophysical neutrinos below the atmospheric neutrino flux, as indicated in the plot (cf.~Fig.~\ref{fig:hese_energy}). The corresponding fit to the eight-year $\nu_\mu+\bar\nu_\mu$ analysis is shown in red.}
\label{fig:fluxes}
\end{figure}

\section{Status Of the Observations of Cosmic Neutrinos}
\label{sec:cosmicnu}

For neutrino astronomy, the first challenge is to select a pure sample of neutrinos, roughly 100,000 per year above a threshold of 0.1\,TeV for IceCube, in a background of ten billion cosmic-ray muons (see Fig.~\ref{fig:earth}), while the second is to identify the small fraction of these neutrinos that is astrophysical in origin, roughly at the level of tens of events per year. Atmospheric neutrinos are an overwhelming background for cosmic neutrinos, at least at neutrino energies below $\sim300$\,TeV. Above this energy, the atmospheric neutrino flux 
reduces to less than one event per year, even in a kilometer-scale detector, and thus events in that energy range are cosmic in origin.

\begin{figure}[t]\centering
\includegraphics[width=0.7\linewidth]{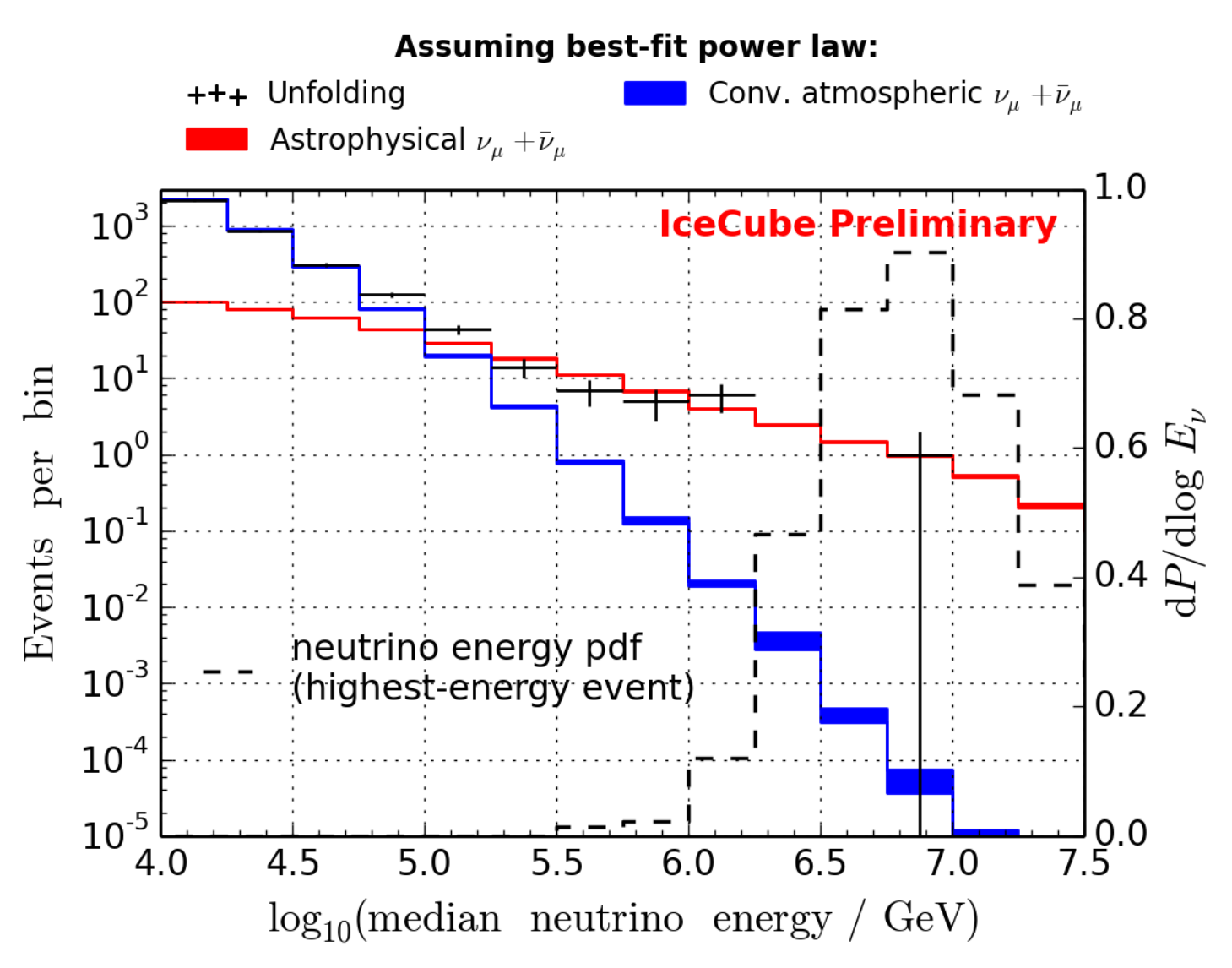}
\caption[]{Distribution of tracks initiated by muon neutrinos that have traversed the Earth, i.e., with zenith angle less than $5^\circ$ above the horizon, as a function of the neutrino energy. Due to the event-by-event variation of the energy transferred to and lost by the muon before it reaches the detector, the initial neutrino energy is shown by its median. The black crosses represent the data. The blue colored band shows the expectation for the atmospheric neutrino flux, while the red line represents a power-law fit to the cosmic component with spectral index $\Gamma=2.13$. Additionally, the probability density function for the neutrino energy of the highest energy event is shown assuming the best-fit spectrum (dashed line).}\label{fig:diffusenumu}
\end{figure}

Using the Earth as a filter, a flux of neutrinos has been identified that is predominantly of atmospheric origin. IceCube has measured this flux over three orders of magnitude in energy with a result that is consistent with theoretical calculations. However, with eight years of data, an excess of events is observed at energies beyond 100\,TeV~\cite{Aartsen:2015rwa,Aartsen:2016xlq,Aartsen:2017mau}, which cannot be accommodated by the atmospheric flux; see Fig.~\ref{fig:fluxes}. Allowing for large uncertainties on the extrapolation of the atmospheric component to higher energy, the statistical significance of the excess astrophysical flux is $6.7\sigma$. While IceCube measures only the energy of the secondary muon inside the detector, from Standard Model physics we can infer the energy spectrum of the parent neutrinos. The best-fit neutrino spectrum then allows deriving the probability distribution of neutrino energy for individual events. For instance, for the highest energy event, shown in Fig.~\ref{fig:erniekloppo} on the right, the median energy of the parent neutrino is about 7\,PeV as indicated in Fig.~\ref{fig:diffusenumu}. Note that this calculation~\cite{Aartsen:2016xlq} takes into account the additional tracks from charged current interactions of $\nu_\tau+\bar\nu_\tau$ as well as resonant interactions of $\bar\nu_e$ with electrons (Glashow resonance) assuming a democratic composition of neutrino and antineutrino flavors. Independent of any calculation, the energy lost by the muon inside the instrumented detector volume is $2.6\pm0.3$\,PeV. The cosmic neutrino flux is well described by a power law with a spectral index $\Gamma=2.19\pm0.10$ and a normalization at 100\,TeV neutrino energy of $(1.01^{+0.26}_{-0.23})\,\times10^{-18}\,\rm GeV^{-1}\rm cm^{-2} \rm sr^{-1}$~\cite{Aartsen:2017mau}. The error range is estimated from a profile likelihood using Wilks' theorem and includes both statistical and systematic uncertainties. The neutrino energies contributing to this power-law fit cover the range from 119\,TeV to 4.8\,PeV.

\begin{figure}[t]\centering
\includegraphics[width=0.95\textwidth]{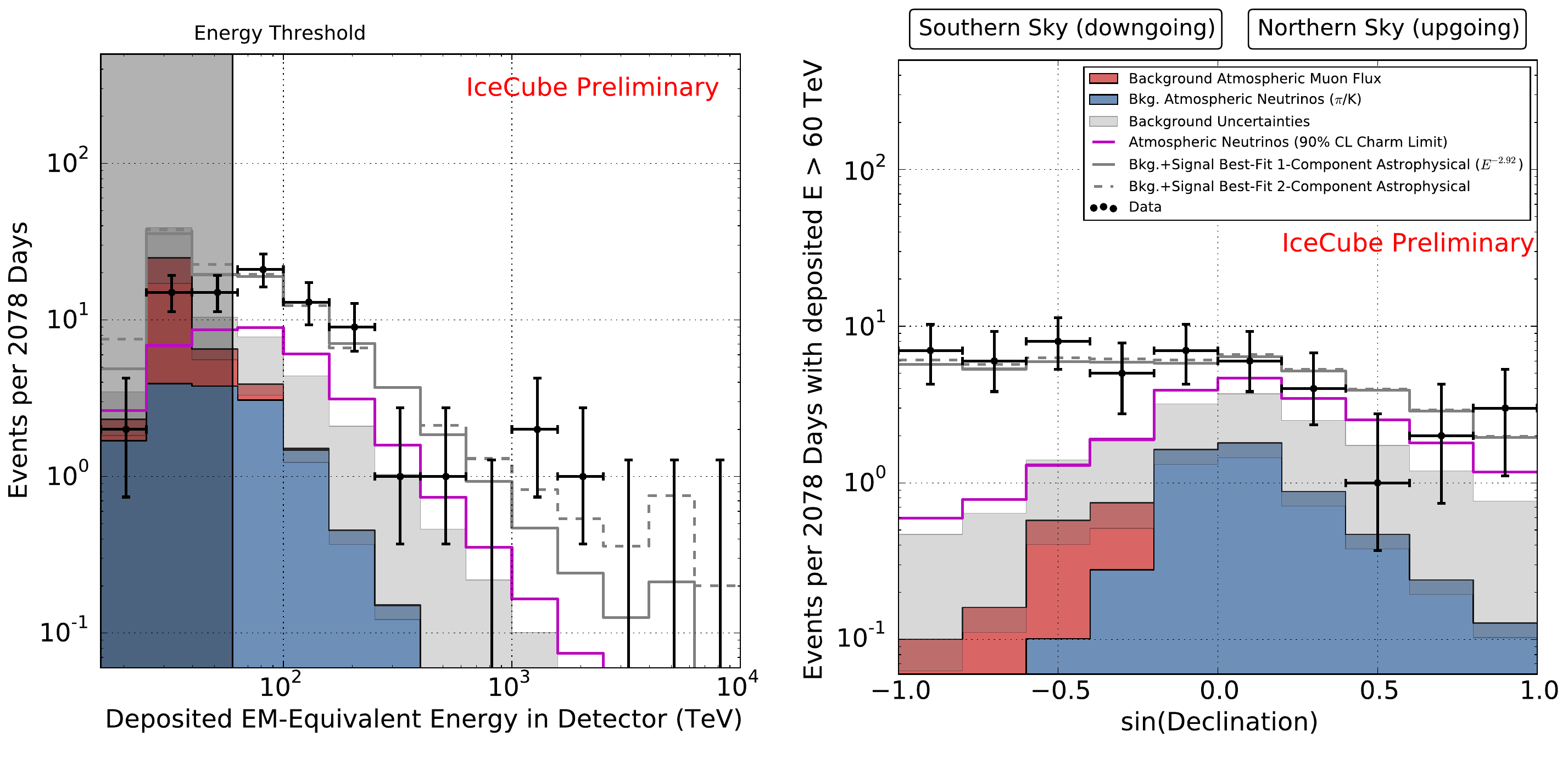}
\caption[]{{\bf Left Panel:} Deposited energies, by neutrinos interacting inside IceCube, observed in six years of data~\cite{Aartsen:2017mau}. The gray region shows uncertainties on the sum of all backgrounds. The atmospheric muon flux (blue) and its uncertainty is computed from simulation to overcome statistical limitations in our background measurement and scaled to match the total measured background rate. The atmospheric neutrino flux is derived from previous measurements of both the $\pi, K$, and charm components of the atmospheric spectrum \protect\cite{Aartsen:2013vca}. Also shown are two fits to the spectrum, assuming a simple power-law (solid gray) and a broken power-law (dashed gray). {\bf Right Panel:} The same data and models, but now showing the distribution of events with deposited energy above 60~TeV in declination. At the South Pole, the declination angle $\delta$ is equivalent to the distribution in zenith angle $\theta$ related by the identity, $\delta = \theta-\pi/2$. It is clearly visible that the data is flat in the Southern Hemisphere, as expected from the contribution of an isotropic astrophysical flux.}
\label{fig:hese_energy}
\end{figure}
 
However, using only two years of data, it was the alternative HESE method, which selects neutrinos interacting inside the detector, that revealed the first evidence for cosmic neutrinos~\cite{Aartsen:2013bka,Aartsen:2013jdh}. The segmentation of the detector into a surrounding veto and active signal region has been optimized to reduce the background of atmospheric muons and neutrinos to a handful of events per year, while keeping most of the cosmic signal. Neutrinos of atmospheric and cosmic origin can be separated not only  by using their well-measured energy but also on the basis that background atmospheric neutrinos reaching us from the Southern Hemisphere can be removed because they are accompanied by particles generated in the same air shower where they originate. A sample event with a light pool of roughly one hundred thousand photoelectrons extending over more than 500 meters is shown in the left panel of Fig.~\ref{fig:erniekloppo}. With PeV energy, and no trace of accompanying muons from an atmospheric shower, these events are highly unlikely to be of atmospheric origin. It is indeed important to realize that the muon produced in the same pion or kaon decay as an atmospheric neutrino will reach the detector provided that the neutrino energy is sufficiently high and the zenith angle sufficiently small~\cite{Schonert:2008is,Gaisser:2014bja}. As a consequence, PeV atmospheric neutrinos originating from above the detector have a built-in self-veto in the HESE analysis by their accompanying atmospheric muons.

\begin{figure}[t]\centering
\includegraphics[width=0.7\linewidth]{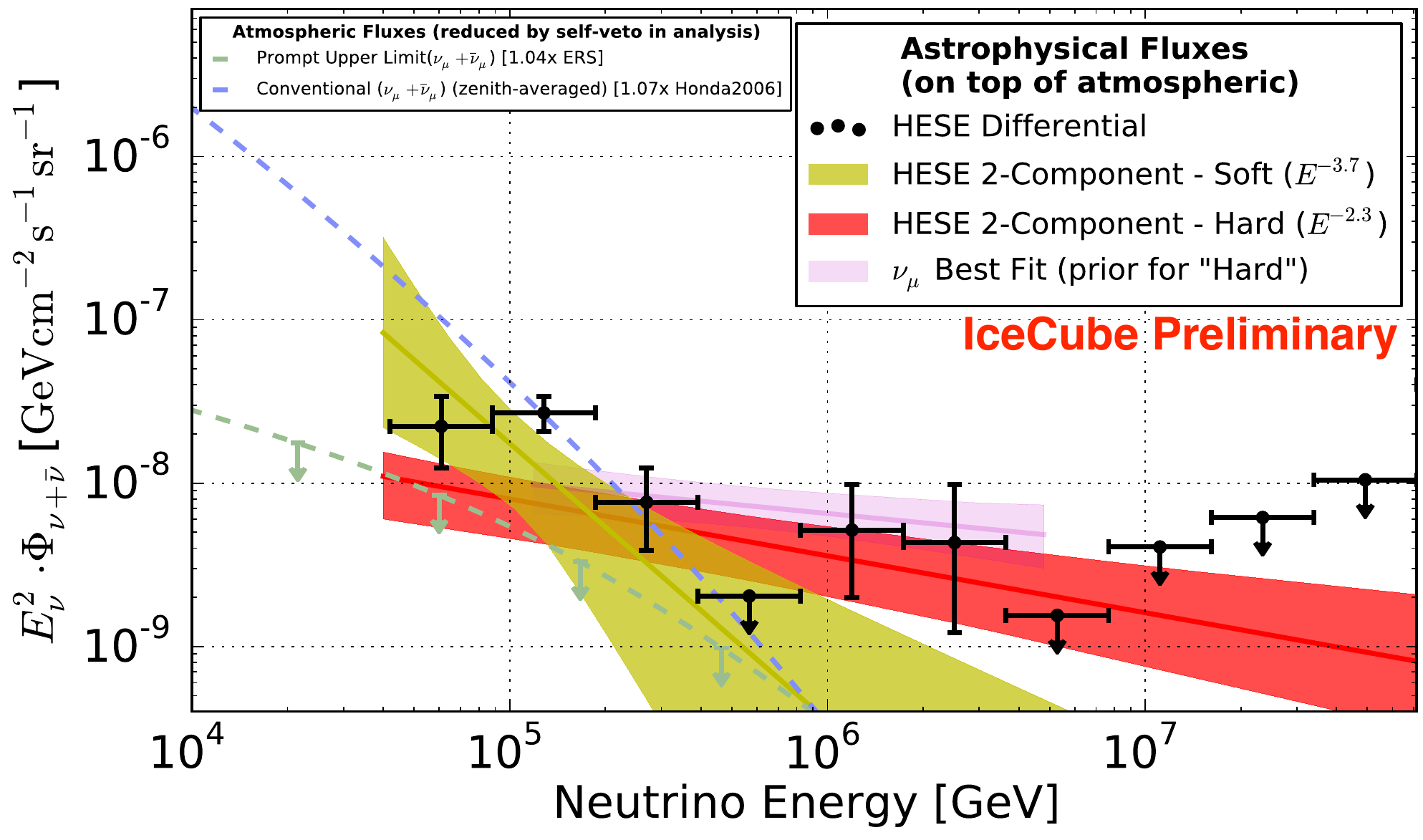}
\caption[]{Unfolded spectrum for six years of HESE neutrino events starting inside the detector. The yellow and red bands show the $1\sigma$ uncertainties on the result of a two-power-law fit. Superimposed is the best fit to eight years of the upgoing muon neutrino data (pink). Note the consistency of the red and pink bands. Figure from Ref.~\cite{Aartsen:2017mau}.}
\label{fig:two_component}
\end{figure}

The deposited energy and zenith dependence of the high-energy starting events collected in six years of data~\cite{Aartsen:2017mau} is compared to the atmospheric background in Fig.~\ref{fig:hese_energy}. The expected number of events for the best-fit astrophysical neutrino spectrum following a two-component power-law fit is shown as dashed lines in the two panels. The corresponding neutrino spectrum is also shown in Fig.~\ref{fig:two_component}. It is, above an energy of $200$\,TeV, consistent with a power-law flux of muon neutrinos penetrating the Earth inferred by the data shown in Fig.~\ref{fig:diffusenumu}. A purely atmospheric explanation of the observation is excluded at $8\sigma$. In summary, IceCube has observed cosmic neutrinos using both methods for rejecting background. Based on different methods for reconstruction and energy measurement, their results agree, pointing at extragalactic sources whose flux has equilibrated in the three flavors after propagation over cosmic distances~\cite{Aartsen:2015ivb} with $\nu_e:\nu_\mu:\nu_\tau \sim 1:1:1$.

The six-year data set contains a total of 82 neutrino events with deposited energies ranging from 60\,TeV to 10\,PeV. The data in both Fig.~\ref{fig:diffusenumu} and Fig.~\ref{fig:hese_energy} are consistent with an astrophysical component with a spectrum close to $E^{-2}$ above an energy of $\sim 200$\,TeV. An extrapolation of this high-energy flux to lower energy suggests an excess of events in the $30-100$\,TeV energy range over and above a single power-law fit; see Fig.~\ref{fig:two_component}. This conclusion is supported by a subsequent analysis that has lowered the threshold of the starting-event analysis~\cite{Aartsen:2016tpb} and by a variety of other analyses. The astrophysical flux measured by IceCube is not featureless; either the spectrum of cosmic accelerators cannot be described by a single power law or a second component of cosmic neutrino sources emerges in the spectrum. Because of the self-veto of atmospheric neutrinos in the HESE analysis, i.e.,~the veto triggered by accompanying atmospheric muons, it is very difficult to accommodate the component below 100\,TeV as a feature in the atmospheric background.

\begin{figure}[t]\centering
\includegraphics[width=0.95\linewidth,viewport=5 30 645 360,clip=true]{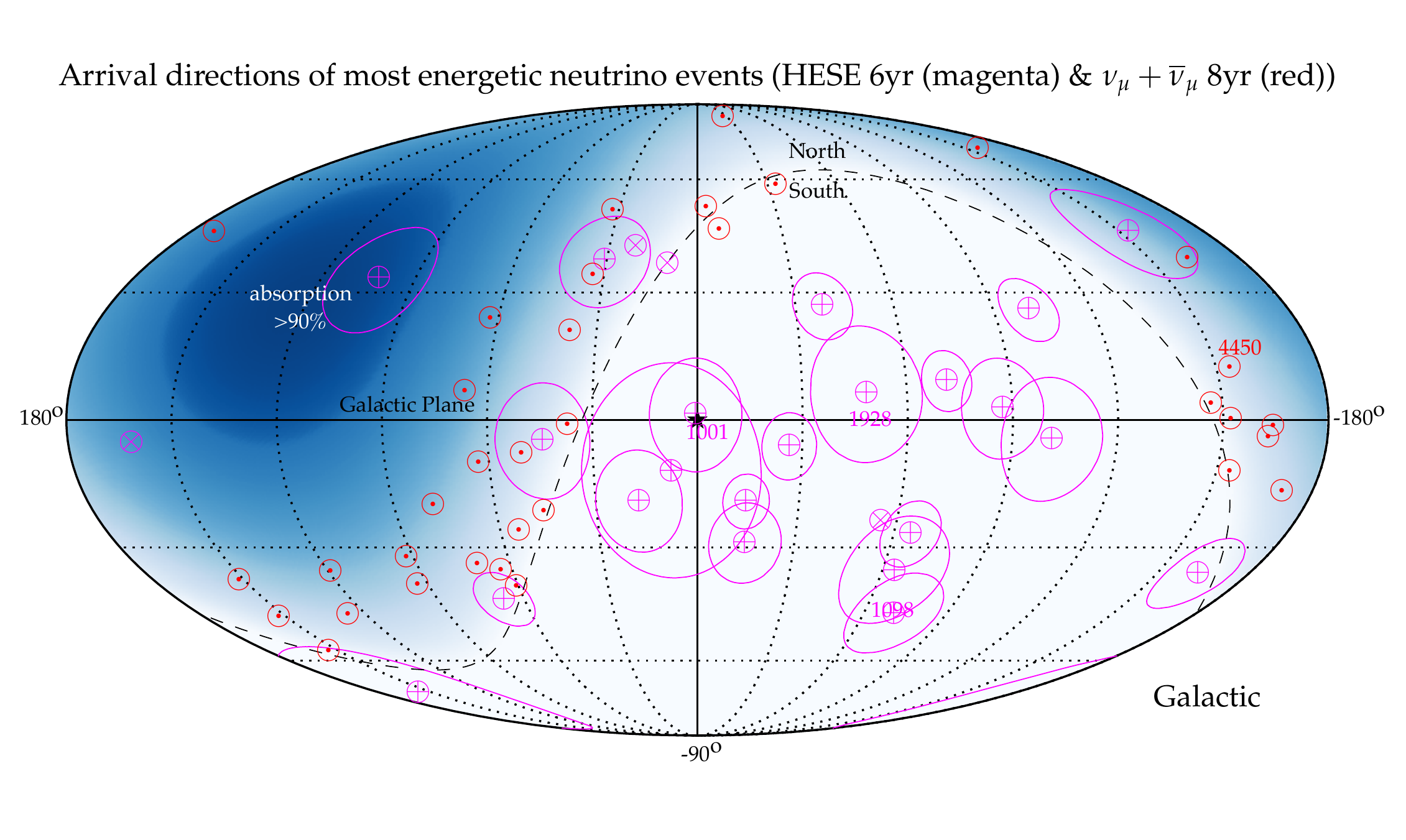}
\caption[]{Mollweide projection in Galactic coordinates of the arrival direction of neutrino events. We show the results of the eight-year upgoing track analysis~\cite{Aartsen:2017mau} with reconstructed muon energy $E_\mu\gtrsim200$~TeV ({$\odot$}). The events of the six-year high-energy starting event (HESE) analysis with deposited energy larger than 100\,TeV (tracks {$\otimes$} and cascades {$\oplus$}) are also shown~\cite{Aartsen:2014gkd,Aartsen:2015zva,Aartsen:2017mau}. The thin circles indicate the median angular resolution of the cascade events ({$\oplus$}). The blue-shaded region indicates the zenith-dependent range where Earth absorption of 100~TeV neutrinos becomes important, reaching more than 90\% close to the nadir. 
The dashed line indicates the horizon and the star ({\scriptsize $\star$}) the Galactic Center. We highlight the four most energetic events in both analyses by their deposited energy (magenta numbers) and reconstructed muon energy (red number).}
\label{fig:NeutrinoMap}
\end{figure}

In Figure \ref{fig:NeutrinoMap} we show the arrival directions of the most energetic events in the eight-year upgoing $\nu_\mu+\bar\nu_\mu$ analysis ($\odot$) and the six-year HESE data sets. The HESE data are separated into tracks ($\otimes$) and cascades ($\oplus$). The median angular resolution of the cascade events is indicated by thin circles around the best-fit position. The most energetic muons with energy $E_\mu>200$~TeV in the upgoing $\nu_\mu+\bar\nu_\mu$ data set accumulate near the horizon in the Northern Hemisphere. Elsewhere, muon neutrinos are increasingly absorbed in the Earth before reaching the vicinity of the detector because of their relatively large high-energy cross sections. This causes the apparent anisotropy of the events in the Northern Hemisphere. Also HESE events with deposited energy of $E_{\rm dep}>100$~TeV suffer from absorption in the Earth and are therefore mostly detected when originating in the Southern Hemisphere. After correcting for absorption, the arrival directions of cosmic neutrinos are isotropic,  suggesting extragalactic sources. In fact, no correlation of the arrival directions of the highest energy events, shown in Fig.~\ref{fig:NeutrinoMap}, with potential sources or source classes has reached the level of $3\sigma$~\cite{Aartsen:2016tpb}.

The absence of strong anisotropies in the arrival direction of IceCube data disfavors scenarios with strong Galactic emission. However, the limited number of events and the low angular resolution of cascade-dominated samples can hide this type of emission. At a minimum, it is possible that some of the data originates in Galactic sources. Various Galactic scenarios have been considered for the diffuse neutrino flux in the TeV-PeV energy range, including the diffuse emission from Galactic CRs~\cite{Ahlers:2013xia,Joshi:2013aua,Neronov:2013lza,Kachelriess:2014oma,Gaggero:2015xza,Neronov:2015osa}, the joint emission of Galactic CR sources~\cite{Fox:2013oza,Gonzalez-Garcia:2013iha,Anchordoqui:2014rca}, or very extended emission from the {\it Fermi bubbles}~\cite{Razzaque:2013uoa,Ahlers:2013xia,Lunardini:2013gva} or the Galactic halo~\cite{Taylor:2014hya,Kalashev:2016euk}. More exotic scenarios consider dark matter decay~\cite{Feldstein:2013kka,Esmaili:2013gha,Bai:2013nga,Murase:2015gea,Boucenna:2015tra,Chianese:2016kpu,Cohen:2016uyg} in the Galactic dark matter halo. Most of these scenarios also predict the production of pionic PeV gamma rays. These gamma rays are absorbed via pair production in the scattering off CMB photons with an absorption length of about 10~kpc. Therefore, the observation of PeV gamma rays would be a ``smoking gun'' of Galactic PeV neutrino emission~\cite{Gupta:2013xfa,Ahlers:2013xia}.

However, the isotropic arrival direction of neutrinos would be a natural consequence of extragalactic source populations. A plethora of models have been considered, including galaxies with intense star formation~\cite{Loeb:2006tw,Murase:2013rfa,He:2013cqa,Anchordoqui:2014yva,Chang:2014hua,Tamborra:2014xia,Bechtol:2015uqb,Senno:2015tra}, cores of active galactic nuclei (AGNs)~\cite{Stecker:1991vm,Stecker:2013fxa,Kalashev:2014vya}, low-luminosity AGNs~\cite{Bai:2014kba,Kimura:2014jba}, quasar-driven outflows~\cite{Wang:2016vbf}, blazars~\cite{Tavecchio:2014eia,Padovani:2014bha,Dermer:2014vaa,Petropoulou:2015upa,Padovani:2015mba,Kadler:2016ygj,Padovani:2016wwn,Neronov:2016ksj}, low-power gamma-ray bursts (GRBs)~\cite{Waxman:1997ti,Ando:2005xi,Murase:2013ffa,Tamborra:2015fzv}, choked GRBs~\cite{Meszaros:2001ms,Senno:2015tsn,Denton:2017jwk}, cannonball GRBs~\cite{Dado:2014mea}, intergalactic shocks~\cite{Kashiyama:2014rza}, galaxy clusters~\cite{Berezinsky:1996wx,Murase:2008yt,Murase:2013rfa,Zandanel:2014pva}, tidal disruption events~\cite{Wang:2015mmh,Senno:2016bso,Dai:2016gtz,Lunardini:2016xwi,Biehl:2017hnb}, or cosmogenic neutrinos~\cite{Roulet:2012rv,Yacobi:2015kga}. We will discuss in the following how we can use multimessenger information to pinpoint the true origin of the neutrino emission.

\section{Multimessenger Interfaces}\label{sec:multimessenger}

\begin{figure}[t]
\centering
\includegraphics[width=0.95\linewidth]{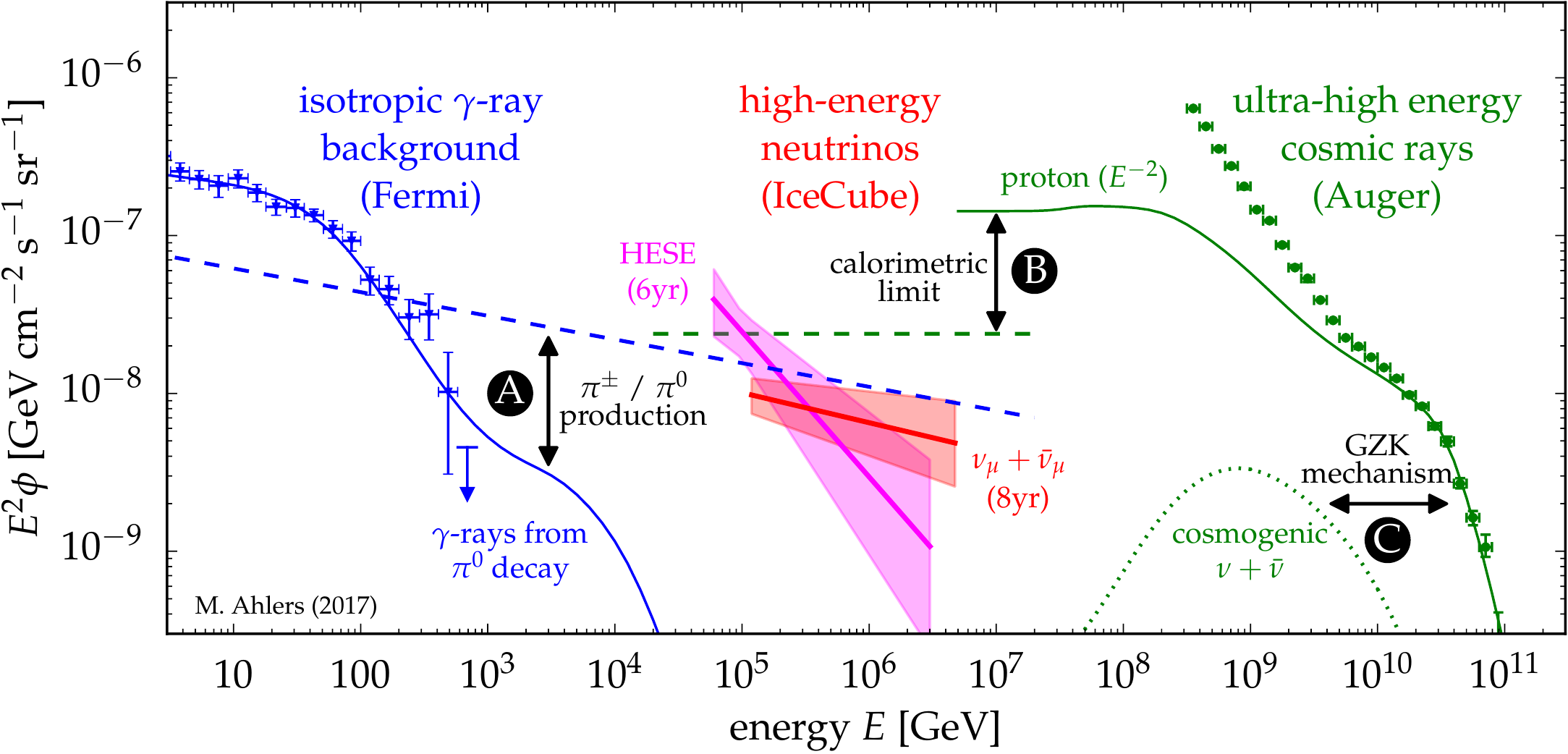}
\caption[]{The spectral flux ($\phi$) of neutrinos inferred from the eight-year upgoing track analysis (red fit) and the six-year HESE analysis (magenta fit) compared to the flux of unresolved extragalactic $\gamma$-ray sources~\cite{Ackermann:2014usa} (blue data) and ultra-high-energy cosmic rays~\cite{Aab:2015bza} (green data). The neutrino spectra are indicated by the best-fit power-law (solid line) and $1\sigma$ uncertainty range (shaded range). We highlight the various multimessenger interfaces: {\bf A:} The joined {production of charged pions ($\pi^\pm$) and neutral pions ($\pi^0$)} in cosmic-ray interactions leads to the emission of neutrinos (dashed blue) and $\gamma$-rays (solid blue), respectively. {\bf B:} Cosmic ray emission models (solid green) of the most energetic cosmic rays imply a maximal flux ({calorimetric limit}) of neutrinos from the same sources (green dashed). {\bf C:} The same cosmic ray model predicts the emission of cosmogenic neutrinos from the collision with cosmic background photons ({GZK mechanism}).}
\label{fig:panorama}
\end{figure}

The most important message emerging from the IceCube measurements is not apparent yet: the prominent and surprisingly important role of protons relative to electrons in the nonthermal universe. To illustrate this point, we show in Fig.~\ref{fig:panorama} the observed neutrino flux $\phi$ in terms of the product $E^2\phi$, which is a measure of its energy density. One can see that the cosmic energy density of high-energy neutrinos is comparable to that of $\gamma$-rays observed with the Fermi satellite~\cite{Ackermann:2014usa} (blue data) and to that of ultra-high-energy (UHE) cosmic rays (above $10^{9}$~GeV) observed, e.g., by the Auger observatory~\cite{Aab:2015bza} (green data). This might indicate a common origin of the signal and provides excellent conditions for multi-messenger studies.
 
A challenge to most galactic and extragalactic scenarios is the large neutrino flux in the range of $10-100$~TeV, which implies an equally high intensity of gamma rays from the decay of neutral pions produced along with the charged pions that are the source of the observed neutrino flux~\cite{Ahlers:2015lln}. For extragalactic scenarios, this gamma-ray emission is not directly observed because of strong absorption of photons by $e^+e^-$ pair production in the extragalactic background light (EBL) and CMB. 
The high-energy leptons initiate electromagnetic showers of repeated inverse-Compton scattering and pair production in the CMB that eventually yield photons that contribute to the Fermi $\gamma$-ray observations in the GeV-TeV range.

The extragalactic $\gamma$-ray background observed by Fermi~\cite{Ackermann:2014usa} has contributions from  identified point-like sources on top of an isotropic $\gamma$-ray background (IGRB) shown in Fig.~\ref{fig:panorama}. This IGRB is expected to consist mostly of emission from the same class of $\gamma$-ray sources that are individually below Fermi's point-source detection threshold (see, e.g., Ref.~\cite{DiMauro:2015tfa}). A significant contribution of $\gamma$-rays associated with IceCube's neutrino observation would have the somewhat surprising implication that indeed many extragalactic $\gamma$-ray sources are also neutrino emitters, while none has been detected so far.

Another intriguing observation is that the high-energy neutrinos observed at IceCube could be related to the sources of UHE CRs. The simple argument is as follows: UHE CR sources can be embedded in environments that act as ``storage rooms'' for cosmic rays with energies far below the ``ankle'' ($E_{\rm CR}\ll1$EeV). This energy-dependent trapping can be achieved via cosmic ray diffusion in magnetic fields. While these cosmic rays are trapped, they can produce $\gamma$-rays and neutrinos via collisions with gas. If the conditions are right, this mechanism can be so efficient that the total energy stored in low-energy cosmic rays is converted to that of $\gamma$-rays and neutrinos. These ``calorimetric'' conditions can be achieved in starburst galaxies~\cite{Loeb:2006tw} or galaxy clusters~\cite{Berezinsky:1996wx}. We will discuss these multimessenger relations in more detail next.

\subsection{IceCube Neutrinos and Fermi Photons}

Photons are produced in association with neutrinos when accelerated cosmic rays produce neutral and charged pions in interactions with photons or nuclei. Targets include strong radiation fields that may be associated with the accelerator as well as concentrations of matter or molecular clouds in their vicinity. Additionally, pions can be produced in the interaction of cosmic rays with the EBL when propagating through the interstellar or intergalactic background. A high-energy flux of neutrinos is produced in the subsequent decay of charged pions via $\pi^+\to\mu^++\nu_\mu$ followed by $\mu^+ \to e^++\nu_e+\bar\nu_\mu$ and the charge-conjugate processes. High-energy gamma rays result from the decay of neutral pions, $\pi^0\to\gamma+\gamma$. Pionic gamma rays and neutrinos carry, on average, 1/2 and 1/4 of the energy of the parent pion, respectively. With these approximations, the neutrino production rate $Q_{\nu_\alpha}$ (units of ${\rm GeV}^{-1} {\rm s}^{-1}$) can be related to the one for charged pions as
\begin{equation}\label{eq:PIONtoNU}
\sum_{\alpha}E_\nu Q_{\nu_\alpha}(E_\nu) \simeq 3\left[E_\pi Q_{\pi^\pm}(E_\pi)\right]_{E_\pi \simeq 4E_\nu}\,.
\end{equation}
Similarly, the production rate of pionic gamma-rays is related to the one for neutral pions as
\begin{equation}\label{eq:PIONtoGAMMA}
E_\gamma Q_{\gamma}(E_\gamma) \simeq 2\left[E_\pi Q_{\pi^0}(E_\pi)\right]_{E_\pi \simeq 2E_\gamma}\,.
\end{equation}
Note, that the relative production rates of pionic gamma rays and neutrinos only depend on the ratio of charged-to-neutral pions produced in cosmic-ray interactions, denoted by $K_\pi = N_{\pi^\pm}/N_{\pi^0}$. Pion production of cosmic rays in interactions with photons can proceed resonantly in the processes $p + \gamma \rightarrow \Delta^+ \rightarrow \pi^0 + p$ and $p + \gamma \rightarrow \Delta^+ \rightarrow \pi^+ + n$. These channels produce charged and neutral pions with probabilities 2/3 and 1/3, respectively. However, the additional contribution of nonresonant pion production changes this ratio to approximately 1/2 and 1/2. In contrast, cosmic rays interacting with matter, e.g.,~hydrogen in the Galactic disk, produce equal numbers of pions of all three charges: $p+p \rightarrow N_\pi\,[\,\pi^{0}+\pi^{+} +\pi^{-}]+X$, where $N_\pi$ is the pion multiplicity. From above arguments we have $K_\pi\simeq2$ for cosmic ray interactions with gas ($pp$) and $K_\pi\simeq1$ for interactions with photons ($p\gamma$). 

With this approximation we can combine Eqs.~(\ref{eq:PIONtoNU}) and (\ref{eq:PIONtoGAMMA}) to derive a simple relation between the pionic gamma-ray and neutrino production rates:
\begin{equation}\label{eq:GAMMAtoNU}
\frac{1}{3}\sum_{\alpha}E^2_\nu Q_{\nu_\alpha}(E_\nu) \simeq \frac{K_\pi}{4}\left[E^2_\gamma Q_\gamma(E_\gamma)\right]_{E_\gamma = 2E_\nu}\,.
\end{equation}
The prefactor $1/4$ accounts for the energy ratio $\langle E_\nu\rangle/\langle E_\gamma\rangle\simeq 1/2$ and the two gamma rays produced in the neutral pion decay. This powerful relation relates pionic neutrinos and gamma rays without any reference to the cosmic ray beam; it simply reflects the fact that a $\pi^0$ produces two $\gamma$ rays for every charged pion producing a $\nu_\mu +\bar\nu_\mu$ pair, which cannot be separated by current experiments.

Before applying this relation to a cosmic accelerator, we have to be aware of the fact that, unlike neutrinos, gamma rays interact with photons of the cosmic microwave background before reaching Earth. The resulting electromagnetic shower subdivides the initial photon energy, resulting in multiple photons in the GeV--TeV energy range by the time the photons reach Earth. Calculating the cascaded gamma-ray flux accompanying IceCube neutrinos is straightforward~\cite{Protheroe1993,Ahlers:2010fw}.

As an illustration, we show a model of $\gamma$-ray and neutrino emission as blue lines in Fig.~\ref{fig:panorama}. We assume that the underlying $\pi^0$ / $\pi^\pm$ production follows from cosmic-ray interactions with gas in the universe. In this way, the initial emission spectrum of $\gamma$-rays and neutrinos from pion decay is almost identical to the spectrum of cosmic rays (assumed to be a power law, $E^{-\Gamma}$), after accounting for the different normalizations and energy scales. The flux of neutrinos arriving at Earth (blue dashed line) follows this initial CR emission spectrum. However, the observable flux of $\gamma$-rays (blue solid lines) is strongly attenuated above 100~GeV by interactions with extragalactic background photons.

The overall normalization of the emission is chosen in a way that the model does not exceed the isotropic $\gamma$-ray background observed by the Fermi satellite (blue data). This implies an upper limit on the neutrino flux shown as the blue dashed line. Interestingly, the neutrino data shown in Fig.~\ref{fig:panorama} saturates this limit above 100~TeV. Moreover, the HESE data that extends to lower energies is only marginally consistent with the upper bound implied by the model (blue dashed line). This example shows that multi-messenger studies of $\gamma$-ray and neutrino data are powerful tools to study the neutrino production mechanism and to constrain neutrino source models~\cite{Murase:2013rfa}.

The matching energy densities of the extragalactic gamma-ray flux detected by Fermi and the high-energy neutrino flux measured by IceCube suggest that, rather than detecting some exotic sources, it is more likely that IceCube to a large extent observes the same universe astronomers do. Clearly, an extreme universe modeled exclusively on the basis of electromagnetic processes is no longer realistic. The finding implies that a large fraction, possibly most, of the energy in the nonthermal universe originates in hadronic processes, indicating a larger role than previously thought. The high intensity of the neutrino flux below 100~TeV in comparison to the Fermi data might indicate that these sources are even more efficient neutrino than gamma-ray sources~\cite{Murase:2015xka,Bechtol:2015uqb}.

IceCube is developing methods, most promisingly real-time multiwavelength observations in cooperation with astronomical telescopes, to identify the sources and build on the discovery of cosmic neutrinos to launch a new era in astronomy~\cite{Aartsen:2016qbu,Aartsen:2016lmt}. We will return to a coincident observation of a flaring blazar on September 22, 2017, further on.

\subsection{IceCube Neutrinos and Ultra-High-Energy Cosmic Rays}
 
The charged pion production rate $Q_{\pi^\pm}$ is proportional to the density of the cosmic-ray nucleons in the beam that produces the pions, $Q_N$, by a ``bolometric'' proportionality factor $f_\pi\leq 1$. For a target with nucleon density $n$ and extension $\ell$, the efficiency factor for producing pions is $f_\pi \simeq 1-\exp(-\kappa\ell\sigma n)$, where the cross section $\sigma$ and inelasticity, i.e., average relative energy loss of the leading nucleon, refer to either $p\gamma$ or $pp$ interactions. The pion production efficiency $f_\pi$ normalizes the conversion of cosmic-ray energy into pion energy on the target as:
\begin{equation}\label{eq:CRtoPION}
E_\pi^2Q_{\pi^\pm}(E_\pi) \simeq f_\pi\, \frac{K_\pi}{1+K_\pi}\,\left[E^2_NQ_N(E_N)\right]_{E_N = E_\pi/\kappa_\pi}\,.
\end{equation}
We already introduced the pion ratio $K_\pi$ in the previous section, with $K_\pi\simeq2$ for $pp$ and $K_\pi\simeq1$ for $p\gamma$ interactions. The factor $\kappa_\pi$ denotes the average inelasticity {\it per pion} that depends on the average pion multiplicity $N_\pi$. For, both, $pp$ and $p\gamma$ interactions this can be approximated as $\kappa_\pi = \kappa/N_\pi \simeq 0.2$. The average energy per pion is then $\langle E_\pi\rangle  = \kappa_\pi E_N$ and the average energy of the pionic leptons relative to the nucleon is $\langle E_\nu\rangle \simeq \langle E_\pi\rangle /4 = (\kappa_\pi/4)E_N \simeq 0.05 E_N$.

In general, the CR nucleon emission rate, $Q_N$, depends on the composition of UHE CRs and can be related to the spectra\footnote{Note that the integrated number of nucleons is linear to mass number, $\int {\rm d}E Q_N(E) = \sum_AA\int {\rm d}EQ_A(E)$.} of nuclei with mass number $A$ as $Q_N(E_N) = \sum_A A^2Q_A(AE_N)$. In the following we will derive a upper limit on diffuse neutrino fluxes under the assumption that UHE CRs are dominated by protons~\cite{Waxman:1998yy,Bahcall:1999yr}. The local emission rate {\it density}, $\mathcal{Q} = \rho_0Q$,  at these energies is insensitive to the luminosity evolution of sources at high redshift and can be estimated to be at the level of $\left[E_p^2\mathcal{Q}_p(E_p)\right]_{10^{19.5}{\rm eV}} \sim (0.5-2.0)\times10^{44} {\rm erg}/{\rm Mpc}^{3}/{\rm yr}$~\cite{Ahlers:2012rz,Katz:2013ooa,Waxman:2015ues}. Note, that CR composition measurements indicate that the mass composition above the ankle also requires a contribution of heavier nuclei. However, the estimated local UHE CR power density based on proton models is a good proxy for that of UHE CR models including heavy nuclei, as long as the spectral index is close to $\Gamma\simeq2$. For instance, a recent analysis of Auger~\cite{Aab:2016zth} provides a solution with spectral index $\Gamma\simeq2.04$ and a combined nucleon density of $[E_N^2\mathcal{Q}_N(E_N)]_{10^{19.5}{\rm eV}} \sim 2.2\times10^{43}\,{\rm erg}/{\rm Mpc}^{3}/{\rm yr}$. 

For the calculation of the (quasi-)diffuse neutrino spectra we start from the contribution of individual sources. A neutrino point-source (PS) at redshift\,$z$ with spectral emission rate $Q_{\nu_\alpha}$ contributes a flux (in units ${\rm GeV}^{-1} {\rm cm}^{-2} {\rm s}^{-1}$ and summed over flavors)
\begin{equation}\label{eq:PS}
\phi^{\rm PS}_{\nu}(E_\nu) = \frac{(1+z)^2}{4\pi d^2_L(z)}\sum_\alpha Q_{\nu_\alpha}((1+z)E_\nu)\,,
\end{equation}
where $d_L$ is the luminosity distance. For the standard $\Lambda$CDM cosmological model~\cite{Olive:2016xmw}, this is simply given by the redshift integral
\begin{equation}
d_L(z) = (1+z)\int\limits_0^z\frac{{\rm d}z'}{H(z')}\,.
\end{equation} 
Here, the Hubble parameter $H$ has a local value of $c/H_0 \simeq 4.4$~Gpc and scales with redshift as $H^2(z) = H^2_0[(1+z)^3\Omega_{\rm m} + \Omega_\Lambda]$, with $\Omega_{\rm m} \simeq 0.3$ and $\Omega_\Lambda\simeq 0.7$. Note that the extra factor $(1+z)^2$ appearing in Eq.~(\ref{eq:PS}) follows from the definition of the luminosity distance and accounts for the relation of the energy flux to the differential neutrino flux $\phi$. The diffuse neutrino flux from extragalactic sources is given by the integral over co-moving volume ${\rm d}V_c = 4\pi (d_L/(1+z))^2 {\rm d}z/H(z)$. Weighting each neutrino source by its density per co-moving volume $\rho(z)$ gives (see, e.g., Ref.~\cite{Ahlers:2014ioa}):
\begin{equation}\label{eq:Jtot}
\phi_{\nu}(E_\nu) = \frac{c}{4\pi}\int_0^\infty\frac{{\rm d}z}{H(z)}\mathcal{\rho}(z)\sum_\alpha Q_{\nu_\alpha}((1+z)E_\nu)\,.
\end{equation} 
In the following, we will assume that the neutrino emission rate $Q_{\nu_\alpha}$ follows a power law $E^{-\Gamma}$. The flavor-averaged neutrino flux can then be written as
\begin{equation}\label{eq:Lnu}
\frac{1}{3}\sum_\alpha E_\nu^2\phi_{\nu_\alpha}(E_\nu) = \frac{c}{4\pi}\frac{\xi_z}{H_0}\frac{1}{3}\sum_\alpha E_\nu^2 \mathcal{Q}_{\nu_\alpha}(E_\nu)\,,
\end{equation}
where $\mathcal{Q}_{\nu_\alpha}= \rho_0Q_{\nu_\alpha}$ is the neutrino emission rate {\it density} and we introduced the redshift factor
\begin{equation}\label{xi}
\xi_z = \int_0^\infty{\rm d}z\frac{(1+z)^{-\Gamma}}{\sqrt{\Omega_\Lambda+(1+z)^3\Omega_{\rm m}}}\frac{\rho(z)}{\rho_0}\,.
\end{equation}
A spectral index of $\Gamma\simeq2.0$ and no source evolution in the local ($z<2$) universe, $\rho(z)=\rho_0$, yields $\xi_z\simeq0.5$. For sources following the star-formation rate, $\rho(z)=(1+z)^3$ for $z<1.5$ and $\rho(z)=(1+1.5)^3$ for $1.5<z<4$, with the same spectral index yields $\xi_z\simeq 2.6$.

We can now derive the observed diffuse neutrino flux related to the sources of UHE CRs. Combining Eqs.~(\ref{eq:PIONtoNU}) and (\ref{eq:CRtoPION}) to relate the local neutrino emission rate density to the CR nucleon rate density, we arrive at the diffuse (per flavor) neutrino flux via Eq.~(\ref{eq:Lnu}):
\begin{equation}\label{eq:WBbound}
\frac{1}{3}\sum_{\alpha}E_\nu^2\phi_{\nu_\alpha}(E_\nu)
\simeq  3\times10^{-8}f_\pi\left(\frac{\xi_z}{2.6}\right)\left(\frac{[E^2_p\mathcal{Q}_p(E_p)]_{E_p=10^{19.5}{\rm eV}}}{10^{44} \,{\rm erg}/{\rm Mpc}^{3}/{\rm yr}}\right)\frac{\rm GeV}{\rm cm\,s\,sr}\,.
\end{equation}
Here, we have assumed $pp$ interactions with $K_\pi=2$. The calorimetric limit, $f_\pi\to 1$, of Eq.~(\ref{eq:WBbound}) corresponds to the {\it Waxman--Bahcall} (WB) upper limit on neutrino production in UHE CR sources~\cite{Waxman:1998yy,Bahcall:1999yr}. As mentioned in the introduction of this section, these calorimetric conditions can be achieved by a rigidity-dependent cosmic ray ``trapping'' considered, e.g., in starburst galaxies~\cite{Loeb:2006tw} or galaxy clusters~\cite{Berezinsky:1996wx}.

It is intriguing that the observed intensity of diffuse neutrinos is close to the level of the WB bound. A more precise correspondence is illustrated by the green lines in Fig.~\ref{fig:panorama} showing a model of UHECR protons that could account for the most energetic cosmic rays (green data). Note, that the cosmic ray data below $10^{10}$~GeV is not accounted for by this model and must be supplied by additional sources, not discussed here. If we now consider the case that the UHECR sources are embedded in calorimeters, we can derive the maximal neutrino emission (green dashed line) from the low-energy tail of the proton model. Interestingly, the observed neutrino flux saturates this calorimetric limit. It is therefore feasible that UHECRs and neutrinos observed with IceCube have a common origin. If this is the case, the neutrino spectrum beyond 200~TeV should reflect the energy-dependent release of cosmic rays from the calorimeters. Future studies of the neutrino spectrum beyond 1~PeV can provide supporting evidence for CR calorimeter. In particular, the transition to a thin environment ($f_\pi\ll1$), that is a necessary condition of UHE CR emission, implies a break or cutoff in the neutrino spectrum.

Note that the proton model in Fig.~\ref{fig:panorama} also contributes to the flux of EeV neutrinos shown as a dotted green line. Ultra-high energy CRs are strongly attenuated by resonant interactions with background photons, as first pointed out by Greisen, Zatsepin and Kuzmin~\cite{Greisen:1966jv,Zatsepin:1966jv} (GZK). This GZK mechanism is responsible for the suppression of the UHECR proton flux beyond $5\times10^{10}$~GeV (``GZK cutoff'') in Fig.~\ref{fig:panorama} (green solid line) and predicts a detectable flux of cosmogenic neutrinos~\cite{Beresinsky:1969qj} (green dotted line). The proton fraction of UHE CRs at $E\geq10^{19.5}$~eV can be probed by EHE neutrino observatories at the level of 10\% if the (all flavor) EeV sensitivity reaches $E^2\phi\simeq 10^{-9}\,{\rm GeV}/{\rm cm}^{2}/{\rm s}/{\rm sr}$ and if the sources follow the evolution of star formation rate. The low energy tail of the same population of UHE CR protons can be responsible for the observed neutrino emission below 10~PeV assuming a calorimetric environment.

\begin{figure}[t]\centering
\includegraphics[width=0.6\linewidth]{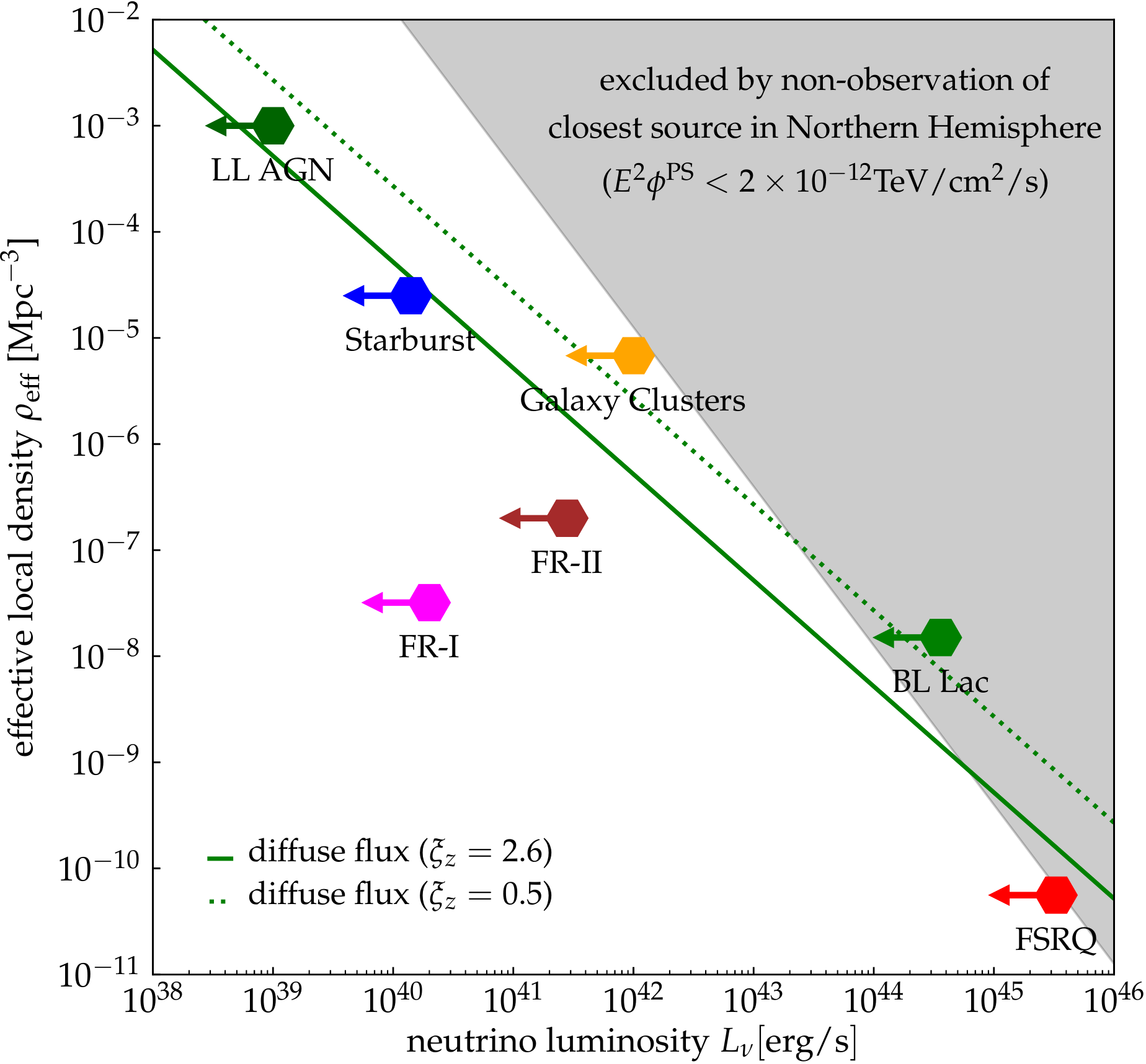}\\[0.3cm]
\caption[]{The effective local density and (maximal) neutrino luminosity of various neutrino source candidates from Ref.~\cite{Mertsch:2016hcd}. The green solid (green dotted) line shows the local density and luminosity of the population of sources responsible for the diffuse neutrino flux of $E^2\phi\simeq 10^{-8}~{\rm GeV}~{\rm cm}^{-2}~{\rm s}^{-1}~{\rm sr}^{-1}$ observed with IceCube, assuming source evolution following the star-formation rate ($\xi_z\simeq2.6$) or no source evolution ($\xi_z\simeq0.5$), respectively. The gray-shaded area indicates source populations that are excluded by the nonobservation of point sources in the Northern Hemisphere ($f_{\rm sky} \simeq 0.5$) with discovery potential $E^2\phi^{\rm PS}\simeq 2\times10^{-12}~{\rm TeV}~{\rm cm}^{-2}~{\rm s}^{-1}$~\cite{Aartsen:2017kru}.
}\label{fig:Luminosity}
\end{figure}

\section{Pinpointing the Astrophysical Sources of Cosmic Neutrinos}
\label{sec:populations}

The flux of neutrinos measured by IceCube provides a constraint on the flux from the individual sources that it is composed of. We can investigate under what circumstances IceCube can detect the neutrino emission from individual, presumably nearby, point sources that contribute to the quasi-diffuse emission. Eq.~(\ref{eq:Lnu}) relates the average luminosity of individual neutrino sources to the diffuse flux that is measured by the experiment to be at the level of $E^2\phi_\nu \simeq 10^{-8}~{\rm GeV} {\rm cm}^{-2} {\rm s}^{-1} {\rm sr}^{-1}$ for energies in excess of $\simeq 100$\,TeV; see~Fig. \ref{fig:fluxes}. From the measurement, we can infer the average emission from a single source~\cite{Ahlers:2014ioa} depending on the local density of sources,
\begin{equation}\label{eq:rate}
\frac{1}{3}\sum_\alpha E_\nu^2 Q_{\nu_\alpha}(E_\nu) \simeq 1.7\times10^{43} \left(\frac{\xi_z}{2.6}\right)^{-1}\left(\frac{\rho_0}{10^{-8}{\rm Mpc}^{-3}}\right)^{-1}{\rm erg}\,{\rm s}^{-1}\,.
\end{equation}
For a homogeneous distribution of sources, we expect, within the partial field of view $f_{\rm sky}$ of the full sky, one source within a distance $d_1$ determined by $f_{\rm sky} 4\pi d_1^3/3\,\rho_0 = 1$. In other words, $d_1$ defines the volume containing one nearby source for a homogeneous source density $\rho_0$. Defining $\phi_1$ as the flux of a source at distance $d_1$, given by Eq.~(\ref{eq:PS}) with $z\simeq0$ and $d_L\simeq d_1$, we can write the probability distribution $p(\phi)$ of finding the {\it closest} source of the population with a flux $\phi$ as (for details see App.~of Ref.~\cite{Ahlers:2014ioa})
\begin{equation}
p(\phi) = \frac{3}{2} \frac{1}{\phi}\left(\frac{\phi_1}{\phi}\right)^{\frac{3}{2}}e^{-\left(\frac{\phi_1}{\phi}\right)^{\frac{3}{2}}}\,.
\end{equation}
The average flux from the closest source is then $\langle \phi\rangle \simeq 2.7\phi_1$, with median $\phi_{\rm med} \simeq 1.3\phi_1$. Applying this to the closest source introduced above, we obtain its per-flavor flux:
\begin{equation}\label{eq:point_diffuse}
\frac{1}{3}\sum_\alpha E_\nu^2\phi_{\nu_\alpha} \simeq 8\times10^{-13}\left(\frac{f_{\rm sky}}{0.5}\right)^{\frac{2}{3}}\left(\frac{\xi_z}{2.6}\right)^{-1}\left(\frac{\rho_0}{10^{-8}{\rm Mpc}^{-3}}\right)^{-\frac{1}{3}} {\rm TeV}\,{\rm cm}^{-2}\,{\rm s}^{-1}\,.
\end{equation}
Interestingly, this value is not far from IceCube's point-source discovery potential at the level of $2\times10^{-12}\,{\rm TeV}\,{\rm cm}^{-2}\,{\rm s}^{-1}$ in the Northern Hemisphere~\cite{Aartsen:2017kru}. 

In Figure~\ref{fig:Luminosity}, we show the local density and luminosity of theorized neutrino sources~\cite{Mertsch:2016hcd}. The gray-shaded region is excluded by the failure to observe these sources as individual point sources, assuming the discovery potential of IceCube in the Northern Hemisphere given in the caption. The green lines show the combination of density and luminosity for sources at the level of the observed IceCube flux, assuming a source density evolution following the star formation rate (solid line) or no evolution (dotted line). We conclude that IceCube is presently sensitive to source populations with local source densities smaller than, conservatively, $10^{-8}\,{\rm Mpc}^{-3}$. Much lower local densities, like BL Lacs FSRQs, are challenged by the nonobservation of individual sources. Some source classes, like Fanaroff-Riley (FR) radio galaxies, have an estimated neutrino luminosity that is likely too low for the observed flux. Note that these estimates depend on the evolution parameter $\xi_z$, and therefore the exact sensitivity estimate depends on the redshift evolution of the source luminosity density. In addition, this simple estimate can be refined by considering not only the closest source of the population but the combined emission of {\it known} local sources; see, e.g.,~Ref.~\cite{Ahlers:2014ioa}. 

The qualitative matching of the energy densities of photons and neutrinos, discussed in the previous section, suggests that the unidentified neutrino sources contributing to the diffuse flux might have already been observed as strong gamma-ray emitters. Theoretical models~\cite{Ajello:2011zi,DiMauro:2013zfa} and recent data analyses~\cite{TheFermi-LAT:2015ykq,Zechlin:2015wdz,Lisanti:2016jub} show that Fermi's extragalactic gamma-ray flux is dominated by blazars. A dedicated IceCube study~\cite{Aartsen:2016lir} of Fermi-observed blazars showed no evidence of neutrino emission from these source candidates. However, the inferred limit on the their quasi-diffuse flux leaves room for a significant contribution to IceCube's diffuse neutrino flux at the 10\% level and increasing towards PeV. This hypothesis is corroborated by the recent observation of the flaring blazar TXS0506+056 in the direction of a very high energy IceCube neutrino.

IceCube detects one well-localized muon neutrino every few minutes as an up-going track event. These events are dominated by low-energy atmospheric neutrinos. IceCube recently installed an automatic filter that selects in real time rare very high energy events that are potentially cosmic in origin and sends the astronomical coordinates to the Gamma-ray Coordinate Network for possible follow-up by astronomical telescopes. The tenth such alert, IceCube-170922A~\cite{2017GCN.21916....1K}, on September 22, 2017, reported a well-reconstructed muon neutrino with a significant probability of originating in outer space rather than in the Earth's atmosphere.

What makes this alert special is that, for the first time, telescopes detected enhanced gamma-ray activity aligned with the cosmic neutrino within less than $0.1^\circ$, a flaring blazar with redshift $z\simeq0.34$~\cite{Paiano:2018qeq}, called TXS0506+056. Originally detected by NASA's Fermi~\cite{2017ATel10791....1T} and Swift\cite{2017ATel10792....1E} satellite telescopes, the alert was followed up by the MAGIC air Cherenkov telescope~\cite{2017ATel10817....1M}. Several other telescopes subsequently observed the flaring blazar. Given where to look, IceCube searched in archival neutrino data, up to and including October 2017, for evidence of neutrino emission at the location of TXS0506+056.  

It is also worth noting that on July 31, 2016, IceCube sent out a similar neutrino alert. The AGILE collaboration, which operates an orbiting X-ray and gamma-ray telescope, reported a day-long blazar flare in the direction of the neutrino, one day before the neutrino detection~\cite{Lucarelli:2017hhh}. Before automatic alerts, in April 2016, the TANAMI collaboration argued for the association of the highest energy IceCube event at the time, dubbed ``Big Bird,'' with the flaring blazar PKS B1424-418~\cite{Kadler:2016ygj}. Finally, AMANDA, IceCube's predecessor, observed three neutrinos in coincidence with a rare flare of the blazar 1ES1959+650, detected by the Whipple telescope in 2002~\cite{Daniel:2005rv}. However, these detections did not reach the significance of the observations triggered by IceCube-170922A.

\section{From Discovery to Astronomy}
\label{sec:future}

Accelerators of CRs produce neutrino fluxes limited in energy to roughly 5\% of the maximal energy of the protons or nuclei. For Galactic neutrino sources, we expect neutrino spectra with a cutoff of a few hundred TeV. Detection of these neutrinos requires optimized sensitivities in the TeV range. At these energies, the atmospheric muon background limits the field of view of neutrino telescopes to the downward hemisphere. With IceCube focusing on high energies, a second kilometer-scale neutrino telescope in the Northern Hemisphere would ideally be optimized to observe the Galactic center and the largest part of the Galactic plane. 

Following the pioneering work of DUMAND~\cite{Babson:1989yy}, several neutrino telescope projects were initiated in the Mediterranean in the 1990s~\cite{Aggouras:2005bg,Aguilar:2006rm,Migneco:2008zz}. In 2008, the construction of the ANTARES detector off the coast of France was completed. With an instrumented volume at about one percent of a cubic kilometer, ANTARES reaches roughly the same sensitivity as AMANDA and is currently the most sensitive observatory for high-energy neutrinos in the Northern Hemisphere. It has demonstrated the feasibility of neutrino detection in the deep sea and has provided a wealth of technical experience and design solutions for deep-sea components. 

While less sensitive than IceCube to a diffuse extragalactic neutrino flux, ANTARES has demonstrated its competitive sensitivity to neutrino emission from the Galactic center~\cite{Adrian-Martinez:2015wey,Adrian-Martinez:2014wzf} and extragalactic neutrino sources in the Southern Hemisphere~\cite{AdrianMartinez:2012rp}. The important synergy between Mediterranean and Antarctic neutrino telescopes has been demonstrated recently by the first joint study of continuous neutrino sources~\cite{Adrian-Martinez:2015ver} as well as neutrino follow-up campaigns of gravitational waves~\cite{ANTARES:2017bia}.

An international collaboration has started construction of a multi-cubic-kilometer neutrino telescope in the Mediterranean Sea, KM3NeT~\cite{Adrian-Martinez:2016fdl}. Major progress has been made in establishing the reliability and the cost-effectiveness of the design. This includes the development of a digital optical module that incorporates 31 3-inch photomultipliers instead of one large photomultiplier tube. The advantages are a tripling of the photocathode area per optical module, a segmentation of the photocathode allowing for a clean identification of coincident Cherenkov photons, some directional sensitivity, and a reduction of the overall number of penetrators and connectors, which are expensive and failure-prone. For all photomultiplier signals exceeding the noise level, time-over-threshold information is digitized and time-stamped by electronic modules housed inside the optical modules. This information is sent via optical fibers to shore, where the data stream will be filtered online for event candidates.

KM3NeT in its second phase~\cite{Adrian-Martinez:2016fdl} will consist of two ARCA units for astrophysical neutrino observations, each consisting of 115 strings (detection units) carrying more than 2,000 optical modules, and one ORCA detector studying fundamental neutrino physics with atmospheric neutrinos. The detection units are anchored to the seabed with deadweights and kept vertical by submerged buoys. The vertical distances between optical modules will be 36\,meters, with horizontal distances between detection units at about 90\,meters. Construction is now ongoing near Capo Passero (east of Sicily).

A parallel effort is underway in Lake Baikal with the construction of the deep underwater neutrino telescope Baikal-GVD (Gigaton Volume Detector)~\cite{Avrorin:2015wba}. The first GVD cluster, named DUBNA, was upgraded in spring 2016 to its final size (288 optical modules, 120 meters in diameter, 525 meters high, and instrumented volume of 6 Mton). Each of the eight strings consists of three sections with 12 optical modules. Deployment of a second cluster was completed in spring 2017.

Further progress requires larger instruments. IceCube therefore proposes as a next step capitalizing on the opportunity of instrumenting $10\rm\,km^3$ of glacial ice at the South Pole and thereby improving on IceCube's sensitive volume by an order of magnitude~\cite{Aartsen:2015dkp}. This large gain is made possible by the unique optical properties of the Antarctic glacier revealed by the construction of IceCube. As a consequence of the extremely long photon absorption lengths in the deep Antarctic ice, the spacing between strings of light sensors can be increased from 125 to over 250 meters without significant loss of performance of the instrument. The instrumented volume can therefore grow by one order of magnitude while keeping the construction budget of a next-generation instrument at the level of the cost of the current IceCube detector. The new facility will increase the event rates of cosmic events from hundreds to thousands over several years.

\section{Conclusion}

IceCube has discovered a flux of extragalactic cosmic neutrinos with an energy density that matches that of extragalactic high-energy photons and UHE CRs. This may suggest that neutrinos and high-energy CRs share a common origin. Identification of the sources by observation of multiple neutrino events from these sources with IceCube will undoubtedly be challenging and require even larger detectors. In the meantime, the possibility exists for revealing the sources by the comprehensive IceCube multimessenger program as illustrated by the observation of IceCube-170922A.

\section*{Acknowledgments}
Discussion with collaborators inside and outside the IceCube Collaboration, too many to be listed, have greatly shaped this presentation. Thanks. 
FH was supported in part by the U.S. National Science Foundation under grants~PLR-1600823 and PHY-1607644 and by the University of Wisconsin Research Committee with funds granted by the Wisconsin Alumni Research Foundation.  MA acknowledges support by Danmarks Grundforskningsfond (grant no.~1041811001) and by a Villum Young Investigator grant (no.~18994) from VILLUM FONDEN.

\section*{References}
\bibliography{references}

\end{document}